\documentclass[conference]{IEEEtran}

\usepackage{fancyhdr}
\usepackage{amsmath,amssymb,amsfonts}
\usepackage[noend]{algcompatible}
\usepackage{graphicx}
\usepackage{textcomp}
\usepackage{color}
\usepackage{hyperref}
\usepackage{xspace}

\usepackage[noend]{algorithm2e}
\usepackage[pdftex]{epsfig}
\usepackage[caption=false]{subfig}  
\usepackage{multicol}
\usepackage{makecell}

\usepackage{url}

\newcommand{\registered}{\textsuperscript{\textregistered}\xspace}
\newcommand{\trademark}{\texttrademark\xspace}

\newcommand{\plc}{\ell^-}

\newcommand{\flc}{\ell^+}
\newcommand{\FLC}{\mathtt{L}^+}
\newcommand{\stpoint}{(\mathbf{r},t)}
\newcommand{\stpprime}{(\mathbf{r}', t')}
\newcommand{\stfield}{X}
\newcommand{\causalfield}{S}

\begin{document}

\title{DisCo: Physics-Based Unsupervised Discovery of Coherent Structures in Spatiotemporal Systems}

%%%%%%%%%%%%%%%%%%%%%%%%%%%%%%%%%%%%%%%%%%%%%%%%%%%%%%%%%%%%%%%%%%%%%%%%%%%%

\author{\IEEEauthorblockN{Anonymous Author(s)}}

 \author{
 \IEEEauthorblockN{Adam Rupe\IEEEauthorrefmark{1}, 
 Nalini Kumar\IEEEauthorrefmark{3}, 
 Vladislav Epifanov\IEEEauthorrefmark{3},
 Karthik Kashinath\IEEEauthorrefmark{2}, 
 Oleksandr Pavlyk\IEEEauthorrefmark{3},\\ 
 Frank Schlimbach\IEEEauthorrefmark{3},  
 Mostofa Patwary\IEEEauthorrefmark{4}, 
 Sergey Maidanov\IEEEauthorrefmark{3},
 Victor Lee\IEEEauthorrefmark{3},
 Prabhat\IEEEauthorrefmark{2},
 James P. Crutchfield\IEEEauthorrefmark{1}\\ 
 \vspace{0.1mm}}
 
\IEEEauthorblockA{\IEEEauthorrefmark{1}Complexity Sciences Center and
 Department of Physics,\\
 University of California at Davis, One Shields Avenue, Davis, CA 95616, USA.}
\IEEEauthorblockA{\IEEEauthorrefmark{3}Intel Corporation, 
 3600 Juliette Ln, Santa Clara, CA 95035, USA. %Email: nalini.kumar@intel.com
}
\IEEEauthorblockA{\IEEEauthorrefmark{2}Lawrence Berkeley National Laboratory, 
 1 Cyclotron Road, M/S 59R4010A, 
 Berkeley, CA 94720, USA
 }
\IEEEauthorblockA{\IEEEauthorrefmark{4}Baidu Research,
 1195 Bordeaux Dr, Sunnyvale, CA 94089, USA.}
 
 %\IEEEauthorblockA{\IEEEauthorrefmark{5} Email: atrupe@ucdavis.edu}
 }

\maketitle

%%%%%%%%%%%%%%%%%%%%%%%%%%%%%%%%%%%%%%%%%%%%%%%%%%%%%%%%%%%%%%%%%%%%%%%%%%%%

\begin{abstract}
%Nalini's edit from SC submission:
Extracting actionable insight from complex unlabeled scientific data is an open challenge and key to unlocking data-driven discovery in science. Complementary and alternative to supervised machine learning approaches, unsupervised physics-based methods based on behavior-driven theories hold great promise. Due to computational limitations, practical application on real-world domain science problems has lagged far behind theoretical development. However, powerful modern supercomputers provide the opportunity to narrow the gap between theory and practical application. We present our first step towards bridging this divide - DisCo - a high-performance distributed workflow for the behavior-driven \emph{local causal state theory}. DisCo provides a scalable unsupervised physics-based representation learning method that decomposes spatiotemporal systems into their structurally relevant components, which are captured by the latent local causal state variables. Complex spatiotemporal systems are generally highly structured and organize around a lower-dimensional skeleton of \emph{coherent structures}, and in several firsts we demonstrate the efficacy of DisCo in capturing such structures from observational and simulated scientific data. To the best of our knowledge, DisCo is also the first application software developed entirely in Python to scale to over 1000 machine nodes, providing good performance along with ensuring domain scientists' productivity. We developed scalable, performant methods optimized for Intel many-core processors that will be upstreamed to open-source Python library packages. Our capstone experiment, using newly developed DisCo workflow and libraries, performs unsupervised spacetime segmentation analysis of CAM5.1 climate simulation data, processing an unprecedented 89.5 TB in 6.6 minutes end-to-end using 1024 Intel Haswell nodes on the Cori supercomputer obtaining 91\% weak-scaling and 64\% strong-scaling efficiency. This enables us to achieve state-of-the-art unsupervised segmentation of coherent spatiotemporal structures in complex fluid flows.

%% Edit from CI paper:
%%Extreme weather is one of the main mechanisms through which climate change will directly impact human society. Coping with such change as a global community requires markedly improved understanding of how global warming drives extreme weather events. While alternative climate scenarios can be simulated using sophisticated models, identifying extreme weather events in these simulations requires automation due to the vast amounts of complex high-dimensional data produced. Atmospheric dynamics, and hydrodynamic flows more generally, are highly structured and largely organize around a lower dimensional skeleton of coherent structures. Indeed, extreme weather events are a special case of more general hydrodynamic coherent structures. We present a scalable physics-based representation learning method that decomposes spatiotemporal systems into their structurally relevant components, which are captured by latent variables known as \emph{local causal states}. For complex fluid flows we show our method is capable of capturing known coherent structures, and with promising segmentation results on CAM5.1 water vapor data we outline the path to extreme weather identification from unlabeled climate model simulation data.

\end{abstract}

%%\keywords{Nonlinear dynamical systems, Python, Unsupervised learning, Clustering, High Performance Computing, Fluid dynamics, Atmosphere}

%%%%%%%%%%%%%%%%%%%%%%%%%%%%%%%%%%%%%%%%%%%%%%%%%%%%%%%%%%%%%%%%%%%%%%%%%%%%

% \fi
\IEEEpeerreviewmaketitle

%%%%%%%%%%%%%%%%%%%%%%%%%%%%%%%%%%%%%%%%%%%%%%%%%%%%%%%%%%%%%%%%%%%%%%%%%%%%

\section{Introduction}

\subsection{Data-Driven Discovery in Science}
Over the last decade, the Data Deluge \cite{Bell1297} has brought dramatic progress across all of science \cite{Chiv18a, Chah17a, Over11a, Sejn14a, Gill16a}. For data-driven science to flourish by extracting meaningful scientific insights \cite{Crut09c, Fagh14a}, new methods are required that discover and mathematically describe complex emergent phenomena, uncover the underlying physical and causal mechanisms, and are better able to predict the occurrence and evolution of these phenomena over time. Increasingly, scientists are leaning upon machine learning (ML) \cite {Mjol01a, Larr06a, Butl18a, Jone17a, Vend18a} and, more recently, deep learning (DL) \cite{Min17a, Carl17a, Reic19a, Bhim18a, Math18a} to fill this role. 

While these techniques show great promise, serious challenges arise when they are applied to scientific problems. 
% To better elucidate the challenges of applying DL methods to science problems, we will focus on a field of study that makes data-driven discovery of utmost and imminent importance - detection and identification of extreme weather events in climate data~\cite{Eman87a,Webs05a} \cite{Webs05a, Wehn10a, Prab12a, Shie18a}. 
To better elucidate the challenges of scientific application of DL methods, we will focus on a particular problem of utmost and imminent importance that necessitates data-driven discovery - detection and identification of extreme weather events in climate data~\cite{Wehn10a, Prab12a, Shie18a}. 
Driven by an ever-warming climate, extreme weather events are changing in frequency and intensity at an unprecedented pace~\cite{Eman87a,Webs05a}. Scientists are simulating a multitude of climate change scenarios using high-resolution, high-fidelity global climate models, producing 100s of TBs of data per simulation. Currently, climate change is assessed in these simulations using summary statistics (e.g. mean global sea surface temperature) which are inadequate for analyzing the full impact of climate change. Due to the sheer size and complexity of these simulated data sets, it is essential to develop robust and automated methods that can provide the deeper insights we seek.

Recently, supervised DL techniques have been applied to address this problem~\cite{mudi17a, jian18a, cohe19d} including one of the 2018 Gordon Bell award winners~\cite{kurt18a}. 
% Further progress, however, has been stymied by two daunting challenges: reliance on labeled training data and interpretability of trained models. 
However, there is an immediate and daunting challenge for these supervised approaches: ground-truth labels do not exist for pixel-level identification of extreme weather events~\cite{Shie18a}.
%Since no ground truth labels exists for pixel-level identification of extreme weather events~\cite{Shie18a}, the DL models used in the above studies are trained using the automated heuristics of TECA~\cite{Prab12a} for proximate labels. 
The DL models used in the above studies are trained using the automated heuristics of TECA~\cite{Prab12a} for proximate labels. 
While the results in \cite{mudi17a} qualitatively show that DL can improve upon TECA, 
the results in \cite{cohe19d} reach accuracy rates over 97\%, essentially reproducing the output of TECA. The supervised learning paradigm of optimizing objective metrics (e.g. training and generalization error)
breaks down here \cite{Fagh14a} since TECA is not ground truth and we do not know how to train a DL model to disagree with TECA in just the right way to get closer to ``ground truth". 

\subsection{Behavior-Driven Theories for Scientific Machine Learning}

% To circumvent these challenges of DL-based approaches, in this paper we present an alternative physics-based unsupervised approach for discovery of coherent structures directly from unlabeled data. Complementary to DL, our approach is grounded in a behavior-driven theory of coherent structures in spatiotemporal systems.  Leveraging physical principles allows behavior-driven models to work with unlabeled data and opens a new line of inquiry into physical interpretability.

With the absence of ground-truth labels, many scientific problems are fundamentally unsupervised problems. Rather than attempt to adapt unsupervised DL approaches to a problem like extreme weather detection, we instead take a behavior-driven approach and start from physical principles to develop a novel physics-based representation learning method for discovering structure in spatiotemporal systems directly from unlabeled data.

At the interface of physics and machine learning, \emph{behavior-driven theories} (e.g. \cite{Will15a, Rung15a,Ye2015a, Vess19a, Zeni19a}) leverage physical principles to extract actionable scientific insight directly from unlabeled data. Focusing directly on system behavior rather than the governing equations is necessitated for complex, nonlinear systems.
For these systems it is generally not possible to deduce properties of emergent behavior from the underlying equations~\cite{Ande72a}. As an example, despite knowing the equations of hydrodynamics and thermodynamics, which critically govern the dynamics of hurricanes, many aspects of how hurricanes form and evolve are still poorly understood~\cite{Eman03a}.

% Due to nonlinearities in the system, deducing properties of emergent behavior from the underlying equations is generally not possible~\cite{Ande72a}. For example, despite knowing the equations of hydrodynamics and thermodynamics, which critically govern the dynamics of hurricanes, many aspects of how hurricanes form and evolve are still poorly understood~\cite{Eman03a}.
% As a response, research on complex, nonlinear systems shifted to focus directly on system behaviors rather than governing equations. Notably, these behavior-driven theories (e.g. \cite{}) leverage physical principles to extract actionable scientific insight directly from unlabeled data and can thus be seen as unsupervised physics-based machine learning methods. 

For the problem of unsupervised segmentation of extreme weather events in climate data, we view these events as particular cases of more general hydrodynamic coherent structures. Atmospheric dynamics, and hydrodynamic flows more generally, are highly structured and largely organize around a lower dimensional skeleton of collective features referred to as \emph{coherent structures}~\cite{Holm12a, Hall15a}. More broadly, coherent structures in spatiotemporal systems can be understood as key organizing features that heavily dictate the dynamics of the full system, and, as with extreme weather, the coherent structures are often the features of interest. 
Project DisCo (`\textbf{Dis}covery of \textbf{Co}herent Structures') combines the behavior-driven local causal state theory of coherent structures with a first-of-its-kind performant and highly scalable HPC implementation in Python.

In Section~\ref{sec:DisCo} we describe the mathematical details of the theory and its use for unsupervised segmentation. In Section~\ref{sec:DisCo} we also present an overview of the distributed DisCo workflow and each of its stages. We then demonstrate its utility by identifying known coherent structures in 2D turbulence simulation data and observational data of Jupiter's clouds from the NASA Cassini spacecraft in Section~\ref{scienceresults}. Finally, we show promising results on CAM5.1 water vapor data and outline the path to extreme weather event segmentation masks. 
%The contributions of our method to the physical interpretability of coherent structures is briefly addressed; a more in-depth discussion of interpretability will appear in a future companion paper. 
% This distributed workflow allowed us to obtain state-of-art unsupervised segmentation results on both observed and simulated scientific data, described in Section~\ref{scienceresults}.

\subsection{Need for High Performance Computing}
Theoretical developments in behavior-driven theories have far outpaced their implementation and application to real science problems due to significant computational demands. Theorists typically use high-productivity languages like Python, which often incur performance penalties, only for prototyping their method and demonstrating its use on small idealized data sets. Since these prototypes aren't typically optimized for production level performance, their use in science applications with big datasets is limited. To solve real science problems, domain scientists often have to rewrite applications, or portions of, in programming languages like C, C++, and Fortran\cite{nerscpython}. %Key stages in the algorithm are memory and I/O intensive, rather than compute bound \cite{Mont15a}, necessitating distributed implementation. 

Making high-productivity languages performant and scalable on HPC systems requires highly optimized platform-specialized libraries with easy-to-use APIs, seamlessly integrated distributed-memory processing modes with popular Python libraries (like scikit-learn), efficient use of JIT compilers like Numba etc. In Project DisCo, we use all these techniques to enable optimized Python code from prototype development to production deployment on more than 1000 nodes of an HPC system. This brings us closer to bridging the performance and productivity disconnect that typically exists in HPC, and streamlining the process from theoretical development to deployment at scale for science applications.

A challenge specific to DisCo is the need for distance-based clustering of lightcone data structures (described in more detail in Sec.~\ref{sec:DisCo}). 
Compared to traditional clustering datasets, lightcones are very high-dimensional objects.
Though lightcone dimensionality depends on reconstruction parameters, even the baseline lower bound of \emph{O}(100) is already very high for typical implementations of clustering methods. To facilitate discovery, our experiments used lightcones with dimension as high as $4495$. 
%To tackle real world problems, we need even larger lightcones, by an order of magnitude or more. 
Also, creation of lightcone vectors increases the on-node data by $O(lightcone\_dimension * 2)$. In our largest run, we process 89.5 TB of lightcone data, which is several orders of magnitude larger than previously reported lightcone-based methods.

To enable novel data-driven discovery at the frontiers of domain science with the ability to process massive amounts of high-dimensional data, we created a highly parallel, distributed-memory, performance optimized implementation of DisCo software including two specialized clustering methods (K-Means \cite{kmeans} and DBSCAN \cite{dbscan}). In keeping with our goal of maintaining scientists' software development productivity, the libraries use standard Python APIs (scikit-learn). These distributed implementations will be up-streamed to benefit the growing community of Python developers. % This distributed DisCo workflow allowed us to obtain state-of-art unsupervised segmentation results on both observed and simulated scientific data, described in Section~\ref{scienceresults}.

% While unsupervised methods like K-Means \cite{kmeans} and DBSCAN \cite{dbscan} do not place theoretical limitations on their use with high-dimensional data, the implementations are typically optimized for data with very small dimensions.
% Performing clustering, especially density-based clustering, in such high dimensions at such data scale has largely been left unexplored until DisCo.
% Moreover, due to the large amount of both raw and lightcone data, DisCo requires this high dimensional clustering to be done over multi-node distributed data sets.
% We developed distributed implementations of K-Means and DBSCAN with similar API calls as scikit-learn and optimized. Our DBSCAN implementation was developed specially for high-dimensional data like the lightcones. In this paper, we use evaluate the good scaling performance with both K-Means and DBSCAN for this large-dataset high-dimensional clustering problem in Project DisCo.

\subsection{Contributions}
Project DisCo makes the following contributions:
\begin{itemize}
\item 
%First distributed-memory implementation of local causal state reconstruction allowing unprecedented data processing capability.
% First distributed-memory implementation of a novel physics-based representation learning method allowing unprecedented capability on large scientific data sets.
First distributed-memory implementation of a novel physics-based representation learning method allowing unprecedented data processing capability on large scientific data sets.
\item Performs unsupervised coherent structure segmentation that qualitatively outperforms state-of-the-art methods for complex realistic fluid flows.
\item Demonstrates good single-node, weak scaling, and strong scaling performance up to 1024 nodes.
\item Distributed implementation of K-Means and DBSCAN clustering methods for high-dimensional data using standard Python APIs.
\item Achieves high performance while maintaining developer productivity by using newly developed optimized Python library functions and efficiently using parallelizing compilers.

\end{itemize}

\section{Related Work}
\label{sec:RelWork}
The basic algorithm for real-valued local causal state reconstruction used by DisCo largely follows that of LICORS \cite{Goer12a, Mont15a}. Without an HPC implementation, LICORS focused on statistical properties of the algorithm, e.g. convergence, and small proof-of-concept experiments. Further, this work used the point-wise entropy over local causal states for coherent structure filters \cite{Shal06a}, but this approach cannot produce objective segmentation masks, as our method is capable of. 
%See \cite{Rupe18b} for further details.   

The first real-valued local causal state reconstruction was done in \cite{Jani07a}, which also analyzed complex fluid flows and climate simulations. They were able to work with these data sets due to efficient data reuse and data sub-sampling from a low-productivity single-node implementation written from scratch. Even with these optimizations in their implementation, DisCo produces much higher resolution results with our high-productivity HPC optimized implementation. Compare the bottom row of Fig.~5 in \cite{Jani07a} with Fig.~\ref{fig:science} in Sec.~\ref{scienceresults}. They also used the local causal state entropy, and so were also not capable of a structural segmentation analysis. 

Lagrangian Coherent Structures (LCS) is a collection of \\approaches grounded in nonlinear dynamical systems theory that seeks to describe the most repelling, attracting, and shearing material surfaces that form the skeletons of Lagrangian particle dynamics \cite{Hall15a}. These approaches are the structural segmentation methods for fluid flows most relevant to DisCo. \cite{Hadj17a} gives a survey of LCS methods, including two benchmark data sets we use here. This provides us a key point of comparison to the state-of-the-art for method validation, given in Section~\ref{scienceresults}.   

DisCo's segmentation semantics are built on a structural decomposition provided by the local causal states. Such a decomposition is similar to empirical dimensionality reduction methods, such as PCA~\cite{Joll11a} and DMD~\cite{Tu14a}. These methods are used extensively in fluid dynamics~\cite{Holm12a} and climate analytics~\cite{Tant15a}.

The key step in the DisCo pipeline requires an unsupervised clustering method. We focus on the popular K-Means~\cite{kmeans} method and the density-based DBSCAN \cite{dbscan}. Further discussion of clustering in the DisCo pipeline is given in Sections \ref{sec:DisCo}, \ref{sec:clustering}, and \ref{scienceresults}.

%There are prototype-based three broad categories of traditional clustering methods (e.g. K-Means), hierarchical clustering methods (e.g., agglomerative clustering), and density-based clustering methods (e.g., DBSCAN).

%K-Means\cite{kmeans} and it's derivative methods are good for identifying spherical clusters and are not theoretically limited by the dimensionality of data. A limitation of the method is that the number of clusters has to be specified apriori. 
%Modern implementations of the K-Means algorithm in Scikit-learn and Intel\registered DAAL \cite{daal} implement K-Means++ \cite{kmeans} which carefully seeds the initial centroids to improve the quality of clustering. 
Several distributed implementations of K-Means have been developed over the years. 
\cite{nwkmeans1} is a C-based implementation that uses both MPI and OpenMP for parallelization. It evenly partitions the data to be clustered among all processes and replicates the cluster centers. At the end of each iteration, global-sum reduction for all cluster centers is performed to generate the new cluster centers. \cite{nwkmeans2} is an extension of this work for larger datasets of billions of points, and \cite{kmeansknc} optimizes K-Means performance on Intel KNC processors by efficient vectorization. The authors of \cite{li2018large} propose a hierarchical scheme for partitioning data based on data flow, centroids(clusters), and dimensions. Our K-Means implementation partitions the problem based on data size, since its application to climate data is a weak scaling problem. We process much larger datasizes, though \cite{li2018large} showcases good performance for much higher dimensionality, up to $O(10E6)$, and clusters $O(10E6)$ than our use case. For comparison, in our capstone problem, in the K-Means stage of DisCo workflow we process $\sim$70E9 lightcones ($\sim$70E6/node) of 84 dimensions into 8 clusters in 2.32 s/iteration on Intel E5-2698 v3 (vs. 2.5E6 samples of 68 dimensions into 10,000 clusters in 2.42 s/iteration on 16 nodes of Intel i7-3770K processors in \cite{li2018large}). We also use a custom distance metric for applying temporal decay (described in Section 3.2.2) which doubles the number of floating point operations.

% \begin{itemize}
%     \item Kmeans on KNC 
%     \item Kmeans with mapreduce 
% \end{itemize}

%The other category of methods for unsupervised clustering is that of density-based methods like DBSCAN \cite{dbscan}. 
%DBSCAN uses a global density threshold to separate regions of different densities into separate non-hierarchical clusters. Density-based methods can identify clusters of arbitrary shapes and sizes. HDBSCAN was designed to find clusters of different densities and create a clustering hierarchy instead of a flat hierarchy \cite{hdbscan}.

Several distributed implementations have been developed for density-based algorithms.  BD-CATS is a distributed DBSCAN implementation using union-find data structures that scales to 8000 nodes on Cori \cite{bdcats}. POPTICS \cite{poptics} is a distributed implementation of the OPTICS algorithm using the disjoint-set data structure and scaled to 3000 nodes. HDBSCAN from Petuum analytics \cite{petuum} uses the NN-Descent algorithm and approximates the k-NN graph. While their method has been shown to work for high-dimensional data, it has not been shown to work at scale. Other implementations such as PDBSCAN \cite{pdbscan}, PSDBSCAN \cite{psdbscan},  PDSDBSCAN \cite{pdsdbscan}, HPDBSCAN \cite{hpdbscan}, etc. have been shown to scale well, but they use specialized indexing structures like k-d trees or ball-trees, which are sub-optimal for clustering high-dimensional data. To the best of our knowledge, this is the first implementation to demonstrate clustering to \emph{O}(100) dimensional data at this data scale (56 million points per node and 57 billion points in total).
\section{Description of the DisCo project}
\label{sec:DisCo}

Project DisCo combines the first distributed HPC implementation of local causal state reconstruction with theoretical advances in using local causal states to decompose spatiotemporal systems into structurally relevant components and objectively identify coherent structures~\cite{Rupe18b, Rupe2018c}. 
When fully deployed, DisCo promises to discover and identify extreme weather events in climate data using an unsupervised segmentation analysis with local causal states. We now outline the mathematical basis for this claim.

\begin{figure}[h] 
\centering
\includegraphics[trim={1.8cm 4.5cm 6cm 0.3cm},clip, width=0.45\textwidth, height=5cm, keepaspectratio]{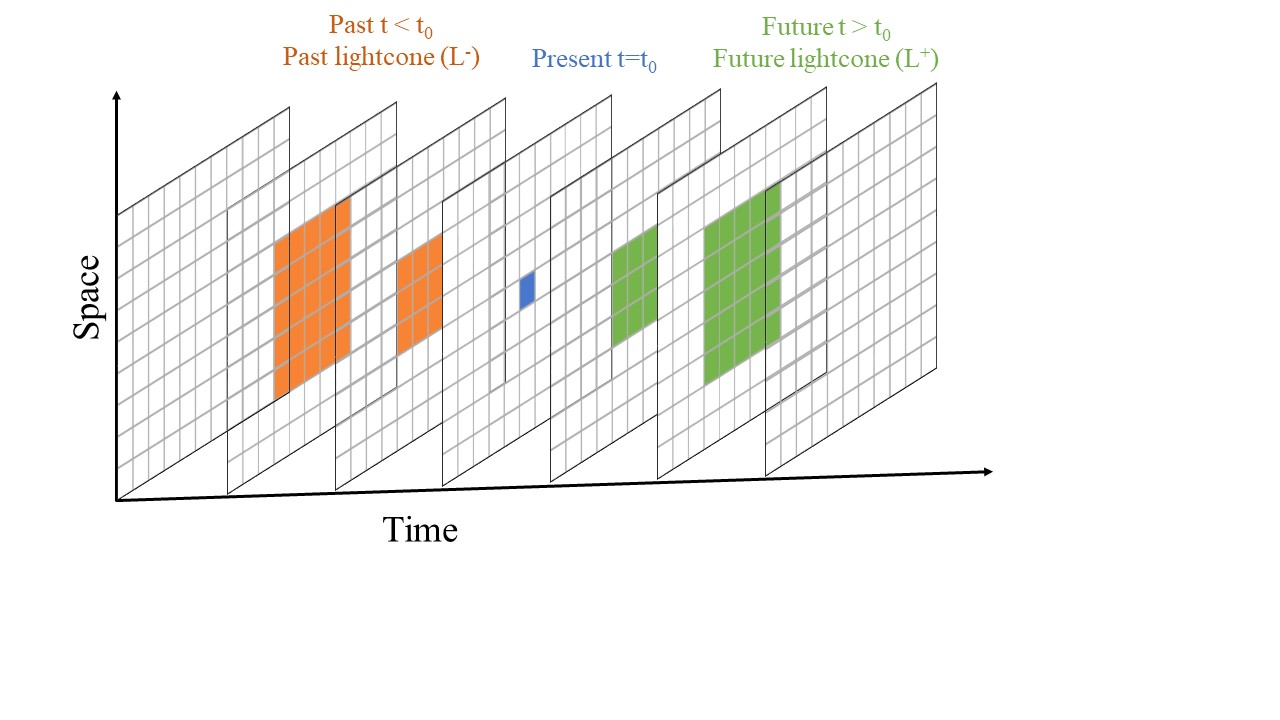}
\caption{\small 2+1D lightcone template with past horizon $h^-=2$, future horizon $h^+=2$, speed of information propagation $c=1$.}
\label{fig:lightcones}	
\end{figure}

\subsection{Local Causal States - Theory}
%Local causal states are a part of a body of behavior-driven theory, known as \emph{computational mechanics} \cite{Crut12a}, that seeks to capture pattern and structure in complex dynamical systems in a principled and constructive manner. Central to this effort is the notion of minimal optimally predictive models, which are uniquely defined through the \emph{causal equivalence relation}
Similar to the intuition behind autoencoder neural networks, pattern and structure of dynamical systems derives from optimal prediction with minimal resources. Thus the mathematical representation of a system's structure is learned through a minimal, optimally predictive stochastic model~\cite{Wolf84a,Gras86,Shal98a}. A body of behavior-driven theory, known as \emph{computational mechanics}~\cite{Crut12a}, gives a principled and constructive realization of this idea through the \emph{causal equivalence relation}
\begin{align*}
\mathrm{past}_i \sim_\epsilon \mathrm{past}_j \iff \Pr(\mathrm{Future} | \mathrm{past}_i) = \Pr(\mathrm{Future} | \mathrm{past}_j).
\end{align*}
Two pasts are causally equivalent if they make the same prediction of the future. The equivalence classes generated by the causal equivalence relation are the \emph{causal states}. They are the unique minimal sufficient statistic of the past for optimally predicting the future \cite{Shal98a}. 
%We emphasize that the causal equivalence relation does not rely on any governing equations of motion, and the resulting model %consisting of the causal states and their state transitions 
%gives a mathematical representation of the dynamical structure of the system behavior. 

Generalizing to spatiotemporal systems, \emph{lightcones} are used as \emph{local} notions of past and future. The past lightcone $\plc$ of a point $\stpoint$ in spacetime field $\stfield$ is defined as the set of all points in the past (up to some finite horizon $h^-$) that could possibly influence $\stfield\stpoint$:
\begin{align*}
% \plc \stpoint \equiv \big\{\stfield \stpprime : \; (t-h^-) \leq t' \leq t \; \mathrm{and} \;
% ||\mathrm{r}' - \mathrm{r}|| \leq c(t - t') \big \}~,
\plc \stpoint \equiv \big\{\stfield \stpprime : \; t^- \leq t' \leq t \; , \;
||\mathrm{r}' - \mathrm{r}|| \leq c(t - t') \big \}~,
\end{align*}
where $c$ is the speed of information propagation in the system and $t^\pm = t \pm h^\pm$. 
%For simplicity, we follow convention and use the Chebyshev distance when constructing lightcones on Cartesian grids. 
The future lightcone $\flc$ of $\stfield\stpoint$ is similarly the set of all points in the future (up to a finite horizon $h^+$) that $\stfield\stpoint$ can possibly affect;
\begin{align*}
% \flc \stpoint \equiv \big\{\stfield \stpprime : \; t < t' \leq (t+h^+) \; \mathrm{and} \;
% ||\mathrm{r}' - \mathrm{r}|| \leq c ( t' - t)\big\} ~.
\flc \stpoint \equiv \big\{\stfield \stpprime : \; t < t' \leq t^+ \; , \;
||\mathrm{r}' - \mathrm{r}|| \leq c ( t' - t)\big\} ~.
\end{align*}

From this we arrive at the \emph{local causal equivalence relation}:
\begin{align*}
% \plc_i \sim_\epsilon \plc_j \iff \Pr(\FLC | \plc_i) = \Pr(\FLC | \plc_j) \iff \epsilon(\plc_i) = \epsilon(\plc_j) ~.
\plc_i \sim_\epsilon \plc_j &\iff \Pr(\FLC | \plc_i) = \Pr(\FLC | \plc_j) \\
\plc_i \sim_\epsilon \plc_j &\iff \epsilon(\plc_i) = \epsilon(\plc_j) ~.
\end{align*}
The associated equivalence classes are the \emph{local causal states} \cite{Shal03a}. They are the unique minimal sufficient statistics of past lightcones for optimal prediction of future lightcones. Each local causal state $\xi$ is the set of all past lightcones with $\Pr(\FLC|\plc) = \Pr(\FLC|\xi)$. The $\epsilon$-function, which generates the causal equivalence classes, maps from past lightcones to local causal states; $\epsilon: \plc \mapsto \xi$. 

Segmentation is achieved by applying the $\epsilon$-function to all points in spacetime, mapping the observable field $\stfield$ to its latent local causal state field $\causalfield = \epsilon(\stfield)$ in a process known as \emph{causal filtering}.
Every feature $x = X(\vec{r},t)$ is mapped to its classification label (local causal state) via its past lightcone $\xi = S(\vec{r}, t) = \epsilon\bigr(\ell^-(\vec{r}, t)\bigl)$.
Crucially, this ensures the latent field $\causalfield$ shares the same coordinate geometry with the observable field $\stfield$ such that $S(\vec{r}, t)$ is the local latent variable corresponding to the local observable $X(\vec{r}, t)$. This means the learned representation is directly utilizable for discovering pattern and structure in the physical observable field. In particular, coherent structure in $\stfield$ are identified through locally broken symmetries in $\causalfield$~\cite{Rupe18b}.

\subsection{Local Causal States - Reconstruction}
%It is these objects, local causal states and the epsilon map, that DisCo reconstructs directly from unlabeled data and uses to perform structural segmentation.
%Below we detail local causal state reconstruction, then outline the distributed DisCo pipeline. 
The core reconstruction parameters are the past lightcone horizon, future lightcone horizon, and speed of information propagation: $(h^-, h^+, c)$. These define the lightcone template, as shown in Figure~\ref{fig:lightcones}. 

\subsubsection{Lightcone extraction}
The main task in local causal state reconstruction is empirical estimation of the conditional distributions, $\Pr(\FLC | \plc)$, known as \emph{future morphs}. Ultimately this comes down to counting past lightcone - future lightcone pairs, $(\plc, \flc)$. Thus the first step is to extract all such lightcone pairs from the given spacetime field(s) $\stfield$ and store them in the paired lists ([\texttt{plcs}], [\texttt{flcs}]). Lightcones are stored as flattened vectors with dimension $ = \sum_{d=0}^{h^\pm} (2dc + 1)^2$.
%Note that there will be points along the boundary of $\stfield$ where full lightcones are not present. The collection of such points is called \emph{the margin} of $\stfield$. 

% \begin{algorithm}
% \SetAlgoLined
% \KwData{spacetime field $\stfield$}
% \KwResult{lists [\texttt{plcs}] and [\texttt{flcs}]}
%     \For{spacetime coordinates $(t,y,x) \in \stfield$}{
%         read values of $\PLC(t,y,x)$ in canonical order\;
%         write values to flattened array $\plc$\;
%         add $\mathbf{\plc}$ to [$\texttt{plcs}$]\;
%         read values of $\FLC(t,y,x)$ in canonical order\;
%         write values to flattened array $\mathbf{\flc}$\;
%         add $\mathbf{\flc}$ to [$\texttt{flcs}$]\;
%         }
%  \caption{Lightcone extraction}
%  \label{alg:extract}
% \end{algorithm}

\subsubsection{Cluster lightcones}
For real-valued systems, like the fluid flows considered here, unique $(\plc, \flc)$ pairs will never repeat. Some form of discretization is needed for counting. The best way to do this is to discretize over the space of lightcones, rather than discretizing the original data itself \cite{Jani07a}. We do this by performing (separate) distance-based clustering on the space of past lightcones and the space of future lightcones~\cite{Goer12a}. 

%To this end we introduce lightcone distance equivalence and the $\gamma$-function. 
Let $\mathbf{C^-}$ be the set of clusters over the real-valued past lightcones that results from some distance-based clustering of [\texttt{plcs}], with individual clusters denoted as $C^-$ and stored in [\texttt{pasts}]. Two lightcones are considered $\gamma$-equivalent if they are assigned to the same distance-based cluster:
\begin{align*}
\ell^-_i \sim_{\gamma} \ell^-_j &\iff \ell^-_i \in C^-_\alpha \; \mathrm{and} \; \ell^-_j \in C^-_\alpha \\
\ell^-_i \sim_{\gamma} \ell^-_j &\iff \gamma(\plc_i) = \gamma(\plc_j)~.
\end{align*}
%In words, two lightcones are $\gamma$-equivalent if they are closely associated in lightcone-space, according to the particular clustering method chosen. 
The $\gamma$-function maps past lightcones to their associated distance-based cluster. 
%Past lightcones and future lightcones are clustered independently. 

All prior work has used Euclidean distance for lightcone clustering. This gives uniform weight to all points within the finite time horizon of the lightcone, and no weight to all points outside. To smooth this step discontinuity, we introduce a \emph{lightcone distance} with an exponential temporal decay. Consider two finite lightcones given as flattened vectors $\mathbf{a}$ and $\mathbf{b}$, each of length n;
\begin{align*}
\mathrm{D}_{\mathrm{lightcone}}(\mathbf{a}, \mathbf{b}) \equiv \sqrt{(a_1 - b_1)^2 + \ldots + \mathrm{e}^{-\tau d(n)}(a_n - b_n)^2}
~,
\end{align*}
where $\tau$ is the temporal decay rate and $d(i)$ is the temporal depth of the lightcone vector at index $i$. 

% \begin{algorithm}
% \SetAlgoLined
% \KwData{lists [\texttt{plcs}] and [\texttt{flcs}]}
% \KwResult{list [\texttt{pasts}] of past lightcone cluster labels and list [\texttt{futures}] of future lightcone cluster labels}
%     \Begin{
%         \textbf{Input}: [\texttt{plcs}]\;
%         perform distance-based clustering using                     $\mathrm{D}_{\mathrm{lightcone}}(\mathbf{\plc_i}, \mathbf{\plc_j})$\; 
%         write cluster assignment labels to [\texttt{pasts}] such that\;
%         each $C^-_i = \;$ [\texttt{pasts}]$_i$ = $\gamma^-(\plc_i)$; $\; \plc_i = \;$ [\texttt{plcs}]$_i$\;
%         \textbf{Output}: [\texttt{pasts}]\;
%         }
        
%     \Begin{
%         \textbf{Input}: [\texttt{flcs}]\;
%         perform distance-based clustering using                     $\mathrm{D}_{\mathrm{lightcone}}(\mathbf{\flc_i}, \mathbf{\flc_j})$\; 
%         write cluster assignment labels to [\texttt{futures}] such that\;
%         each $C^+_i = \;$ [\texttt{futures}]$_i$ = $\gamma^+(\flc_i)$; $\; \flc_i = \;$ [\texttt{flcs}]$_i$\;
%         \textbf{Output}: [\texttt{futures}]\;
%         }
%  \caption{Lightcone clustering}
% \end{algorithm}

\subsubsection{Build morphs}
 After clustering [\texttt{plcs}] and [\texttt{flcs}] to produce [\texttt{pasts}] and [\texttt{futures}], respectively,
 %the real-valued vectors in [\texttt{plcs}] and [\texttt{flcs}] are replaced with cluster assignment labels in [\texttt{pasts}] and [\texttt{futures}] that take values from the finite sets $\mathbf{C}^-$ and $\mathbf{C}^+$. With this 
 we can empirically estimate the future morphs $\Pr(\mathrm{L}^+ | \plc)$ using $\Pr(\mathrm{L}^+ | \ell^-) \approx \Pr(\mathrm{\textbf{C}}^+| \; C^-)$. 
The justification for this is the assumption of \emph{continuous histories} \cite[Assumption 3.1]{Goer12a}: if two past lightcones $\ell^-_i$ and $\ell^-_j$ are very close in lightcone-space, their future morphs $\Pr(\mathrm{L}^+ | \ell^-_i)$ and $\Pr(\mathrm{L}^+ | \ell^-_j)$ must be very similar. Using the $\gamma$-function, we state a more actionable version of this assumption, which is implicitly used, but not formally stated, in Ref. \cite{Goer12a}:
\begin{align*}
\gamma(\ell^-_i) = \gamma(\ell^-_j) \implies \epsilon(\ell^-_i) = \epsilon(\ell^-_j)
~.
%\label{eqn:LICORS}
\end{align*}
%This assumes that two past lightcones placed in the same cluster after distance-based clustering is performed must necessarily be causally equivalent, i.e. have the same future morphs. This is a very practical approximation of the continuous histories assumption. 

The conditional distributions $\Pr(\mathbf{C}^+ | C^-)$ are found as rows of the joint distribution matrix $D$, where $D_{i,j} = \Pr(C^-_i, C^+_j)$. To get $D$ we simply count occurrences of pairs $(C^-, C^+)$ in ([\texttt{pasts}], [\texttt{futures}]). 

% \begin{algorithm}
% \SetAlgoLined
% \KwData{[\texttt{pasts}] and [\texttt{futures}]}
% \KwResult{joint distribution matrix $D$}
%     Initialize $D$ as $N^- = |\mathbf{C}^-|$ by $N^+ = |\mathbf{C}^+|$ array of zeros\;
%     \For{$(C^-, C^+)$ in ([\texttt{pasts}], [\texttt{futures}])}{
%         increment $D_{C^-, C^+}$ by $1$ \;
%         }
%  \caption{Build morphs}
% \end{algorithm}

\subsubsection{Causal equivalence}
With the estimated morphs $\Pr(\mathrm{\mathbf{C}}^+ | C^-)$ in hand we can reconstruct causal equivalence of past clusters. Two past clusters $C^-_i$ and $C^-_j$ are $\psi$-equivalent if they have the same conditional distribution over future clusters:
\begin{align*}
C^-_i \sim_\psi C^-_j &\iff \Pr(\mathrm{\mathbf{C}}^+ | C^-_i) = \Pr(\mathrm{\mathbf{C}}^+ | C^-_j) \\
C^-_i \sim_\psi C^-_j &\iff \psi(C^-_i) = \psi(C^-_j)
~.
\end{align*}
The resulting equivalence classes are the approximated local causal states, and the approximation of $\epsilon(\ell^-)$ is given as: 
\begin{align*}
\epsilon(\ell^-) \approx \psi\bigl(\gamma(\ell^-)\bigr)
~.
%\label{eqn:epsilonapprox}
\end{align*} 
We reconstruct $\psi$-equivalence using hierarchical agglomerative clustering. Distribution similarity $\Pr(\mathbf{C}^+ | C^-_i)$\\ $\approx \Pr(\mathbf{C}^+ | \xi_a)$ is evaluated using a chi-squared test with p-value $0.05$.

% \begin{algorithm}
% \SetAlgoLined
% \KwData{joint distribution $D$}
% \KwResult{approximated local causal states and $\epsilon$-map}
%     Initialize empty list [\texttt{states}] of local causal states\;
%     \For{$\Pr(\mathbf{C}^+ | C^-_i) = D_i$ in $D$}{
%         \For{$\xi_a$ in [\texttt{states}]}{
%             \If{$\Pr(\mathbf{C}^+ | C^-_i) \approx \Pr(\mathbf{C}^+ | \xi_a)$}{
%                 add $C^-_i$ to $\xi_a$; $C^-_i \in \xi_a$\;
%                 update $\psi(C^-_i) = \xi_a$\;
%                 \textbf{break}\;
%                 }
%             }
%         \Else{
%             initialize new state as $\xi_b = \{C^-_i,\}$\; 
%             add $\xi_b$ to [\texttt{states}]\;
%             update $\psi(C^-_i) = \xi_b$
%             }
%     }

%  \caption{Causal equivalence}
% \end{algorithm}

\subsubsection{Causal filter}
Using the approximated $\epsilon$-map we can perform spacetime segmentation of $\stfield$ though causal filtering. The $\gamma$-function has already been applied to produce [\texttt{pasts}] by clustering [\texttt{plcs}]. We then apply the learned $\psi$-function from \textit{causal equivalence} to [\texttt{pasts}] to produce [\texttt{states}]. Because all these lists are in spacetime order, we simply reshape [\texttt{states}] to get the approximated local causal state field $\causalfield \approx \psi\bigl(\gamma(\stfield)\bigr)$.
%%Note that because we are storing [\texttt{pasts}] we have already pre-applied the $\gamma$-map; $C^-_i \in$ [\texttt{pasts}] $= \gamma^-(\plc_i)$ for $\plc_i \in$ [\texttt{plcs}]. Also note that the past lightcones in [\texttt{plcs}], and thus also the cluster assignments in [\texttt{pasts}], are stored in spacetime-order from $\stfield$. 
% \begin{algorithm}
% \SetAlgoLined
% \KwData{spacetime field $\stfield$}
% \KwResult{local causal state field $S = \epsilon(\stfield)$}
%     Initialize empty local causal state field $S$ with same dimensions as $\stfield$\;
%     \For{spacetime coordinates $(t,y,x) \in \stfield$}{
%         read past lightcone: $\plc_i = \PLC(t,y,x)$\;
%         get local causal state: $\xi_a = \psi\bigl(\gamma(\plc_i)\bigr)$\;
%         write local causal state label in $S$: $S(t,y,x) = a$\;
%         }
% \caption{Causal filter}
% \end{algorithm}

\subsection{Distributed Reconstruction Pipeline}
\begin{enumerate}
\item \textit{Data loading}: Stripe the spacetime data so that the spatial fields for each time-step in $\stfield$ are stored individually to allow for parallel I/O. Let \texttt{workset} be the time-steps that each process will extract lightcones from. Because lightcones extend in time, each process must load extra time-steps (\texttt{halos}), $h^-$ at the beginning of \texttt{workset} and $h^+$ at the end. 
%The past \texttt{halo} is $h^-$ extra time-steps at the start of \texttt{workset} and the future \texttt{halo} is $h^+$ extra at the end. 
Each process loads its \texttt{workset + halos} in parallel. 

\item \textit{Lightcone extraction}: The temporal haloing removes any need for communication during lightcone extraction, which proceeds independently for each process. 
    
\item \textit{Communication barrier}:
Ensure all processes have their \\local [\texttt{plcs}] and [\texttt{flcs}] lists before proceeding. 

\item \textit{Cluster lightcones}:
First cluster the past lightcones across all processes. Store the cluster assignments labels locally, in order. Then do the same for future lightcones. 

\item \textit{Build local morphs}:
Each process counts $(C^-, C^+)$ pairs in its local ([\texttt{pasts}], [\texttt{futures}]) to build $D_{\mathrm{local}}$.

\item \textit{Communication barrier}:
Wait for all processes to build $D_{\mathrm{local}}$.

\item \textit{Build global morphs}:
Execute an all-reduce sum of all $D_{\mathrm{local}}$ to yield $D_{\mathrm{global}}$ across all processes.

\item \textit{Causal equivalence}:
Since each process has $D_{\mathrm{global}}$, they can independently reconstruct the approximated local causal states and $\epsilon$-map.

\item \textit{Causal filter}:
Each process independently applies the $\epsilon$-map to their \texttt{workset} to produce $S_{\mathrm{local}}$.

\item \textit{Write output}:
Each process independently saves $S_{\mathrm{local}}$ with time-order labels so that $S = \epsilon(\stfield)$ can be constructed from all $S_{\mathrm{local}}$. 

\end{enumerate}

% \begin{figure}[h]  
% \centering
% \includegraphics[trim={6cm 0cm 6cm 0cm},clip, width=0.5\textwidth,height=7cm, keepaspectratio]{Figures/pipeline.jpg}
% \caption{Distributed reconstruction pipeline}
% \label{fig:lightcones}	
% \end{figure}
\section{Challenges of Lightcone Clustering}
\label{sec:clustering}
The most significant step, both computationally and conceptually, in the DisCo pipeline is the discretization of lightcone-space via distance-based clustering. While there are many choices for distance-based clustering, we focus on two of the most popular clustering algorithms in the scientific community: K-Means \cite{kmeans} and DBSCAN \cite{dbscan}.
%Recall from above that the justification for lightcone clustering comes from the continuous histories assumption -- two past lightcones that are close in lightcone-space must necessarily be causally equivalent. Also consider that the motivation for using local causal states is for structural decomposition.

The use of clustering in a structural decomposition pipeline, along with the need to conform to the continuous histories assumption, 
would seem to favor a density-based method like DBSCAN over a prototype-based method like K-Means. Density-connectivity should ensure nearby lightcones are clustered together, whereas K-Means must return $K$ clusters and therefore may put cuts in lightcone-space that separate nearby past lightcones, violating the continuous histories assumption. 

Because we don't want to put any geometric restrictions on the structures captured by local causal states, the ability of DBSCAN to capture arbitrary cluster shapes seems preferable to K-Means, which only captures convex, isotropic clusters. Furthermore, the restriction to $K$ clusters, as opposed to an arbitrary number of cluster with DBSCAN, puts an upper bound on the number of reconstructed local causal states. To test these hypotheses we experimented with both K-Means and DBSCAN at scale to evaluate their parallel scaling performance and the quality of clustering in the DisCo pipeline on real-world data sets. These experiments and results are discussed in Sections \ref{sec:setup}, \ref{perfresults} and \ref{scienceresults}.

\subsection{Distributed K-Means}
We developed a distributed K-Means implementation which will be upstreamed to daal4py \cite{daal4py}, a Python package similar in usage to scikit-learn. Daal4Py provides a Python interface to a large set of conventional ML algorithms highly tuned for Intel\registered platforms.
In contrast to other distributed frameworks for ML in Python, daal4py uses a strict SPMD approach, and so assumes the input data to be pre-partitioned. All communication within the algorithms is handled under the hood using MPI.

%Single-node:
Our single-node K-Means implementation performs one iteration of the algorithm in the following way: all data points are split into small blocks to be processed in parallel. For each block, distances from all points within the block to all current centroids are computed. Based on these distances, points are reassigned to clusters and each thread computes the partial sums of coordinates for each cluster.
At the end of the iteration the partials sums are reduced from all threads to produce new centroids. We use Intel\registered AVX2 or Intel\registered AVX512 instructions, depending on the hardware platform, for vectorizing distance computations.

%Multi-node:
Our multi-node K-Means implementation follows the same general pattern: on each iteration current centroids are broadcast to all nodes, each node computes the assignments and partial sums of coordinates for each centroid, and then one of the nodes collects all partial sums and produces new centroids. We use MPI4Py for collecting partial sums. We integrate into various methods for finding the initial set of $K$ centroids - first $K$ feature vectors, $K$ random feature vectors, and K-Means++ \cite{kmeans} - provided by Intel\registered DAAL.

\subsection{Distributed DBSCAN}
%Our DBSCAN implementation comes from a pre-release version of Intel\registered DAAL. 
We developed both single-node and multi-node implementations of DBSCAN optimized for use with high-dimensionality lightcone data.

%Single-node:
The single-node DBSCAN implementation computes neighborhoods without using indexing structures, like k-d tree or ball-tree, which are less suitable for high-dimensional data. The overall algorithmic complexity is quadratic in the number of points and linear in feature size (lightcone dimension). Neighborhood computation for blocks of data points is done in parallel without use of pruning techniques. We use Intel\registered AVX2 or Intel\registered AVX512 instructions, depending on the hardware platform, to compute distances between points, giving a 2-2.5x speed-up compared to the non-vectorized version.

%Multi-node:
For multi-node DBSCAN clustering, the first step is geometric re-partitioning of data to gather nearby points on the same node, inspired by the DBSCAN implementation of \cite{bdcats}.
It is performed using the following recursive procedure:
for a group of nodes we choose some dimension, find an approximate median of this dimension from all points currently stored on a node,
split the current group of nodes into two halves (with value of chosen dimension lesser/greater than the median) and reshuffle all points so that each node contains
only points satisfying the property above. 

Next, do geometric re-partitioning recursively for the two resulting halves (groups) of nodes.
Then each node gathers from other nodes any extra points that fall into its bounding box (extended by the epsilon in each direction) similar to \cite{pdsdbscan}.
Using these gathered points the clustering is performed locally on each node (single node implementation) and the results from all the nodes are then merged into a single clustering.

Because we use an approximate value of the median, the geometric partitions can sometimes have imbalanced sizes. This can impact the overall performance since different nodes will complete local clustering at different times and no node can proceed further until every node has finished. Also, the number of extra points for some geometric partitions lying in low and high density regions of the data set may be different, which may also cause some load imbalance among nodes.

%DBSCAN geometric partitions sizes (strong-scaling):
%        128      256     512     1024
%Min     702282	  350287  138313  70824
%Max	 1064628  575663  295132  172377
%Average 873252	  434340  214884  105156
%Median	 871468	  420309  213259  103666

%\vspace{-2mm}
\section{Experimental Setup}
\label{sec:setup}
Here we describe the data sets used for both the science results and scaling measurements. We also describe the HPC system -- Cori -- on which these computations were performed.

\subsection{Description of the Cori System}
\label{sec:cori}
All of our experiments were run on the Cori system at the National Energy Research Scientific Computing Center (NERSC) at Lawrence Berkeley National Laboratory (LBNL). Cori is a Cray XC40 system featuring 2,004 nodes of Intel\registered Xeon\trademark Processor E5-2698 v3 (Haswell) and 9,688 nodes of Intel\registered Xeon Phi\trademark Processor 7250 (KNL). Both Haswell and KNL nodes were used.% for our experiments.

Haswell compute nodes have two 16-core  Haswell processors. Each processor core has a 32 KB private L1 instruction cache, 32 KB private L1 data and a 256 KB private L2 cache.  The 16 cores in each Haswell processor are connected with an on-die interconnect and share a 40-MB shared L3 cache. Each Haswell compute node has 128 GB of DDR4-2133 DRAM.

Each KNL compute node has a single KNL processor with 68 cores (each with 4 simultaneous hardware threads and 32 KB instruction and 32 KB data in L1 cache), 16 GB of on-package MCDRAM, and 96 GB of DDR4-2400 DRAM. Every two cores share an 1MB L2 (with an aggregate of 32MB total). The 68 cores are connected in a 2D mesh network. All measurements on KNL reported in this paper are performed with the MCDRAM in ``cache'' mode (configured as a transparent, direct-mapped cache to the DRAM).

Compute nodes in both the Haswell and KNL partitions are connected via the high-speed Cray Aries interconnect. Cori also has a Sonnexion 2000 Lustre filesystem, which consists of 248 Object Storage Targets (OSTs) and 10,168 disks, giving nearly 30PB of storage and a maximum of 700GB/sec IO performance.

\subsection{Libraries and Environment}
The DisCo application code is written in Python using both open-source and vendor optimized library packages. We use \texttt{Intel\registered Distribution Of Python (IDP) 3.6.8}. IDP incorporates optimized libraries such as Intel\registered MKL and Intel\registered DAAL for machine learning and data analytics operations to improve performance on Intel platforms. We also use \texttt{NumPy (1.16.1), SciPy (1.2.0), Numba (0.42.1), Intel\registered TBB (2019.4), Intel\registered DAAL (2019.3) and Intel\registered MKL (2019.3)} from Intel\registered Anaconda channels. \texttt{MPI4Py (3.0.0)} is built to use the Cray MPI libraries.

For all K-Means experiments, our optimized implementation was built from source with Cray MPI and ICC (18.0.1 20171018). These will be contributed back to Intel\registered daal4py. We compile our DBSCAN implementation with Intel\registered C/C++ Compilers (ICC 18.0.1 20171018) and without the \texttt{-fp-model strict} compiler switch which can impact the vectorization performance. Both K-Means and DBSCAN are linked to Cray MPI binaries as well.

For scaling tests we installed the \texttt{conda} environments on the Lustre filesystem to improve Python package import times for large runs on Cori \cite{nerscpython}. For K-Means experiments, we run the code with 1 MPI process per Haswell socket and limit the number of TBB threads to 32 on a node with \texttt{-m tbb -p 32} flags to the Python interpreter. For DBSCAN experiments we run the code with 1 MPI process per node and 68 tbb threads on KNL, and 1 MPI process per node with 32 threads on Haswell nodes. 

\subsection{Datasets}
Two benchmark data sets: 2D turbulence and clouds of Jupiter, are chosen for validation against a survey of LCS methods from \cite{Hadj17a}, and a simulated climate data set to demonstrate scientific and scaling performance on a real-world scientific application, as in \cite{kurt18a}.  

The Jupiter data is interpolated RGB video taken over a period of 10 days by the NASA Cassini spacecraft and converted to integer grayscale \cite{cassini}. The 2D turbulence data set is the vorticity field from direct numerical solutions of 2-dimensional Navier-Stokes equations using pseudo-spectral methods in a periodic domain \cite{Faraz}. The climate data set, used for scaling experiments, is simulated data of Earth's climate from the 0.25-degree Community Atmospheric Model (CAM5.1) \cite{MFW}. 
%This data set is complex with high-resolution multi-scale multi-physics phenomena. 
Climate variables are stored on an 1152 x 768 spatial grid (float32), with a temporal resolution of 3 hours. Over 100 years of simulation output is available as NetCDF files. Our hero run processed 89.5 TB of lightcone data (obtained from 580 GB of simulated climate data).

\section{Performance Results}
\label{perfresults}

We performed both K-Means and DBSCAN weak-scaling and strong-scaling experiments with the CAM5.1 climate dataset. K-Means experiments are run on Cori Haswell nodes and DBSCAN experiments are run on both Cori Haswell and Cori KNL nodes. The performance of each stage of the pipeline as well as the total time to solution (including synchronization) for an end-to-end single run is measured in seconds.
These measurements capture the end-to-end capability of the system and software, including the single node optimizations, efficiency of the distributed clustering methods, and interconnect subsystems.
%This excludes time to import python packages but includes the time to load data from the filesystem. 
%We don't measure the time to write results to the filesystem. 
%\cite{Jani07a}

\subsection{K-Means performance} 

\subsubsection{Single-node performance}
Table \ref{tab:kmeans} shows the breakdown of execution time of different stages of the DisCo pipeline developed from scratch on one Haswell and KNL node. 

The \textit{data read} stage simply reads the spacetime field data into memory which is then processed in \textit{extract} to generate lightcone vectors. This involves reading spatiotemporal neighbors of each point in the field and flattening them into an n-dimensional vector. These are serial read/write operations that are unavoidable, but the memory access pattern can be optimized. Using Numba decorators for jitting to improve caching and vectorization performance, we obtained a 64x speedup on Haswell and 134x speedup on KNL node resulting in overall speedup of 16.9x on Haswell and 62x on KNL over the baseline implementation inspired by \cite{Jani07a}. For the \textit{cluster\_lc} stage, we compare our optimized K-Means implementation which gives $\sim$20x better performance than stock scikit-learn \cite{d4pysergey}.
The other three stages take only a small fraction of the execution time and have little to gain from directed optimization.
% Add details on flops to byte ratio.

\begin{table}
\caption{\small Single-node performance of the different stages of the DisCo pipeline before and after optimization}
\label{tab:kmeans}
\begin{tabular}{|l | r r r | r r r |}
\hline
Stage & \multicolumn{3}{c}{Haswell, time(s)}  |& \multicolumn{3}{c}{KNL, time(s)} \\
\hline
& \thead{Baseline} & \thead{Opti-\\ mized} & \textbf{\thead{Speed\\up}} & \thead{Baseline} & \thead{Opti-\\ mized} & \textbf{\thead{Speed\\up}} \\
\hline
data read  &     3.32	&	3.29	&	1   &	8.44	&	7.01	&	1.2    \\
extract  &    519.64	&	7.85	&	65	&	4713.47	&	35.13	&	134    \\
cluster\_lc  &     399.63	&	19.03	& 21	&	513.63	&	26.34	&	19    \\
%cluster\_lc  &     20.15	&	19.03	&	1	&	33.683	&	26.34	&	1.2    \\
morphs  &     0.85	&	0.86	&	1    &	6.19	&	6.16	&	1    \\
equivalence  &    0.002	&	0.002	&	1	&	0.56	&	0.02	&	26    \\
causal filter &     0.14	&	0.14	&	0.3	&	0.49	&	0.49	&	1    \\
\hline
{\bf Total} &   923.58	&	31.20	&	\textbf{29.6}	&	5242.78	&	74.93	&	\textbf{70}\\
\hline 
\end{tabular}
\end{table}

\subsubsection{Multi-node scaling}

\begin{figure*}[h]  
\centering
\begin{minipage}{0.99\columnwidth}
  \includegraphics[trim={8cm 1.5cm 6cm 1cm},clip, width=0.99\columnwidth]{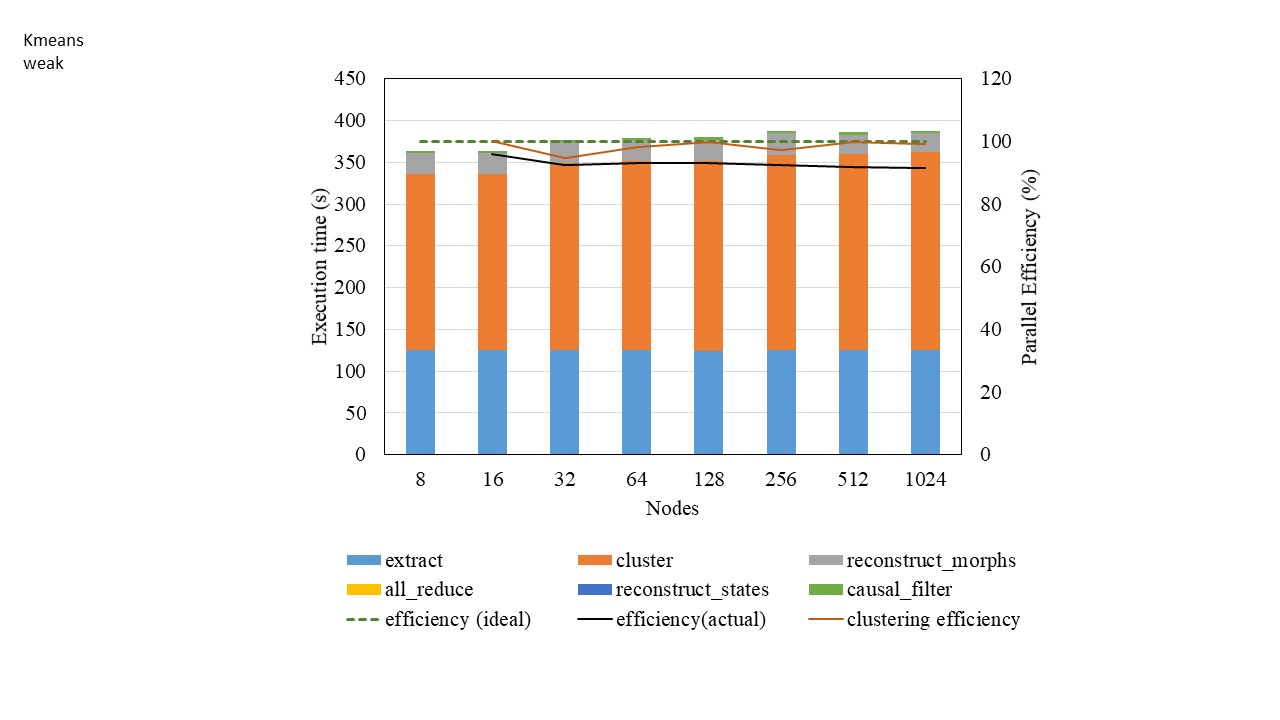}
\end{minipage}%
\begin{minipage}{0.99\columnwidth}
  \includegraphics[trim={7cm 2cm 7cm 1cm},clip, width=0.99\columnwidth]{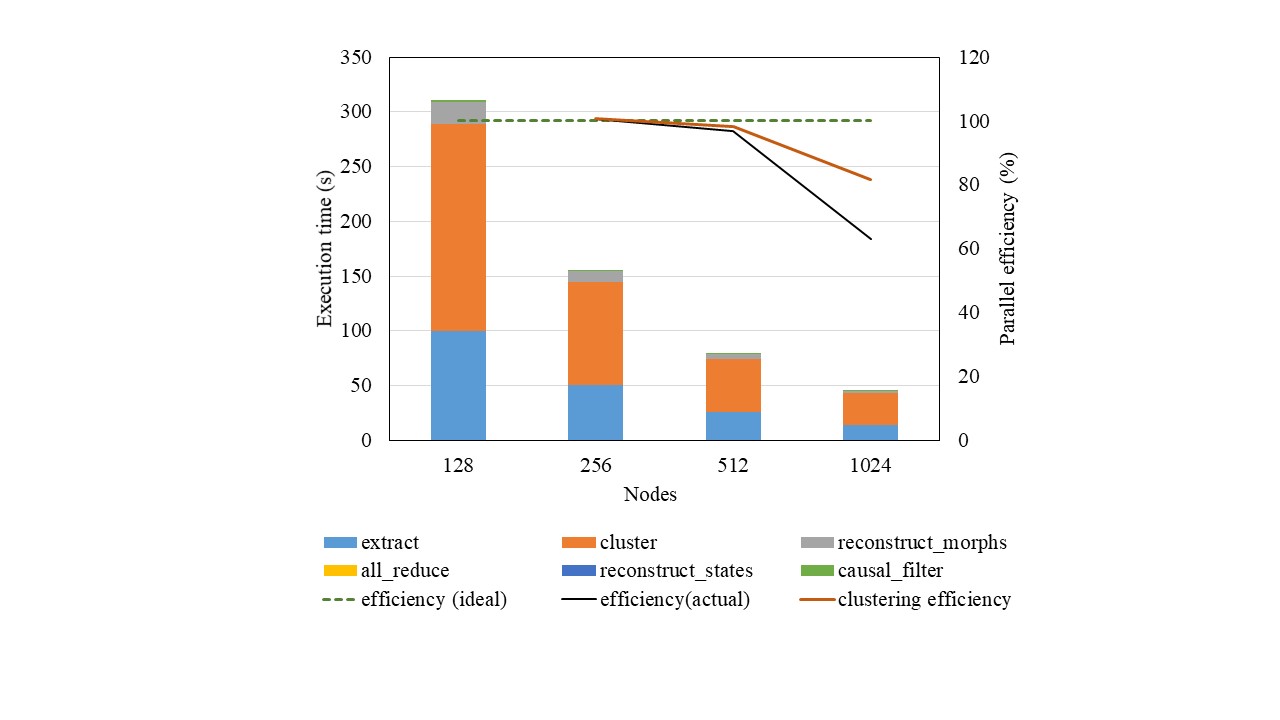}
\end{minipage}
\vspace{-0.25cm}
\caption{\small Breakdown of execution time spent in various stages of the DisCo on Haswell nodes with K-Means. Left : weak scaling and Right: strong scaling. Parallel efficiency are plotted on the secondary axis.}
\label{fig:kmeans-scaling}	
\end{figure*}

All experiments were conducted with $h^{\pm}=3$, $c=1$, 
%\texttt{lightcone depth=3}, \texttt{c=1}, 
number of clusters \texttt{K=8}, and  \texttt{iterations=100} for K-Means clustering. The results are shown in Figure \ref{fig:kmeans-scaling}. \textit{Extract} is embarrassingly parallel and thus, shows excellent scaling. 

For weak scaling on Haswell, we used 220MB/node of raw data (80 timesteps of 1152 x 768 spatial fields). After lightcone extraction (84 dimension vectors of float32 data), the size of input to the clustering stage increases to 87.44GB/node. We achieved weak-scaling efficiency of 91\% at 1024 nodes, measured against performance at 8 nodes. This is expected from increased amounts of time spent in communication at the end of each K-Means iteration as node concurrency increases.

For strong scaling experiments on Haswell, we used 64 timesteps per node on 128 nodes, 32 timesteps per node on 256 nodes, 16 timesteps per node on 512 nodes, and 8 timesteps per node on 1024 nodes. After lightcone extraction the total size of input data to the clustering stage is 54GB. We achieved 64\% overall scaling efficiency and 81\% clustering efficiency at 1024 nodes. At 1024 nodes, the amount of local computation workload on a node is small compared to the number of synchronization steps within K-Means and in the end-to-end pipeline.

\subsection{DBSCAN performance}

\subsubsection{Single-node performance}

\begin{figure*}[ht]  
\centering
\begin{minipage}{0.99\columnwidth}
  \includegraphics[trim={7cm 2.8cm 6cm 1.5cm},clip, width=0.99\columnwidth]{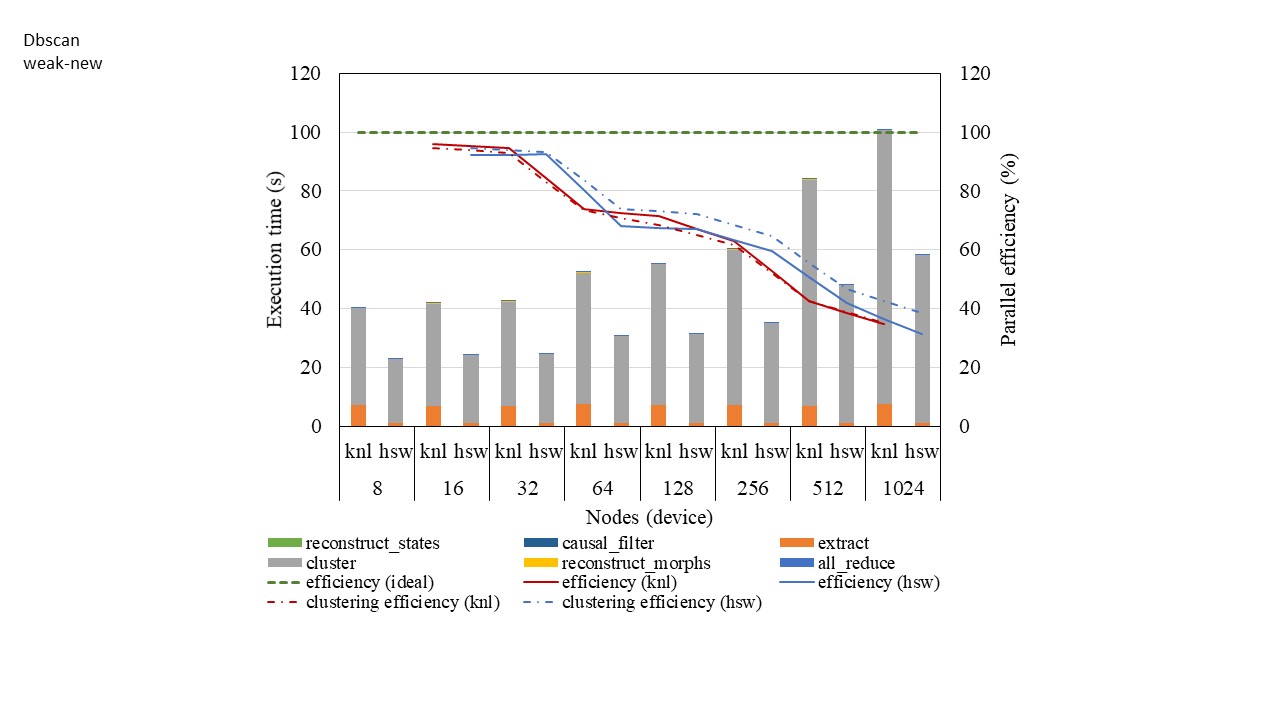}
 \end{minipage}%
\begin{minipage}{0.99\columnwidth}
  \includegraphics[trim={5cm 3cm 8.3cm 1},clip,width=0.99\columnwidth]{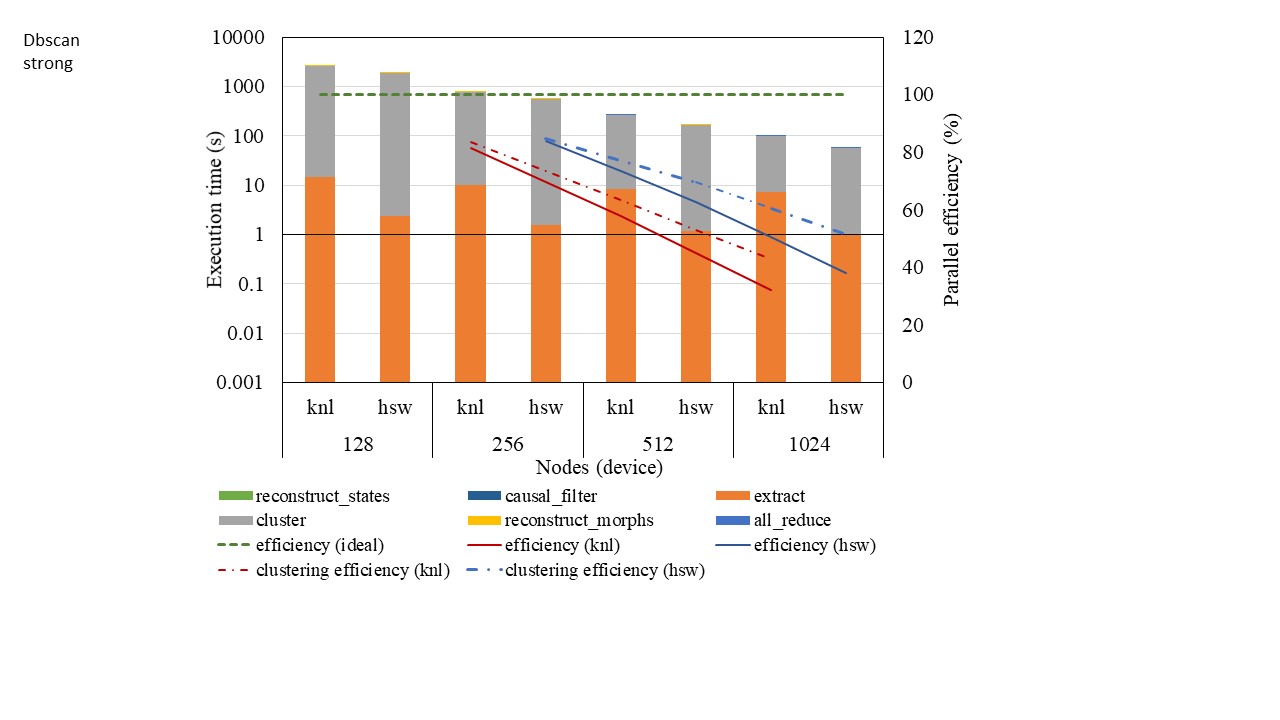}
\end{minipage}
%\vspace{-0.5cm}
  \caption{\small Breakdown of execution time spent in various stages of the DisCo on Haswell and KNL nodes with DBSCAN. Left : weak scaling and Right: strong scaling. Parallel efficiency are plotted on the secondary axis.}
\label{fig:dbscan-scaling}
\end{figure*}

We used the pipeline optimized for K-Means results for which are shown in Table \ref{tab:kmeans}. In the \textit{cluster\_lc} stage, we use our implementation of the DBSCAN algorithm discussed in Section \ref{sec:DisCo}. Designed for high-dimensional clustering, it does not use indexing data structures for nearest neighbor calculations.
On the 2D turbulence data set, the scikit-learn DBSCAN with brute-force neighborhood computation is more than 3x faster than the default scikit-learn DBSCAN, which uses k-d trees, while producing reasonable results (less than $20\%$ noise points). In turn, our DBSCAN implementation is more than 3x faster than the scikit-learn brute implementation (same clustering parameters) due to better on-node parallelization and use of AVX2 and AVX512 vector instructions for computing neighborhoods and distances.
%On one KNL node, with 0.33GB/node of 2D-Turbulence lightcone data, clustering either past or future lightcones took around 25 minutes on KNL and 16 minutes on single-socket Haswell. With 1.5GB/node of 2D-Turbulence lightcone data, clustering past or future lightcones took 11 hours on KNL. The number of lightcones is proportional to the amount of data per node and DBSCAN clustering is quadratic in the number of lightcones.

\subsubsection{Multi-node scaling} 

We performed DBSCAN weak scaling and strong scaling runs using the climate data set on both Haswell and KNL nodes. All experiments were conducted with \texttt{minpts = 10} and \texttt{eps=0.05} for DBSCAN clustering. The results are shown in Figure \ref{fig:dbscan-scaling}. 

For weak scaling on Haswell and KNL, we split a single timestep of the 1152 x 768 spatial field across 8 nodes. At 1024 nodes, we achieved a scaling efficiency of 34.6\%. The poor scaling efficiency can be attributed to several reasons. One, as discussed in Section \ref{sec:DisCo}, distributed DBSCAN uses geometric partitioning to gather adjacent points on the same node. 
Then, at each step, every node clusters its local data subset before merging results among different nodes. Two, since we didn't use indexing data structures to perform local clustering in DBSCAN, the complexity of each step is $O(dimensionality * |size~of~the~partition|^2)$. Third, the total clustering time is equal to the running time of the slowest node, which is the node containing the largest data partition. As the number of nodes increases, it leads to an increase in imbalance in number of points between nodes (\ref{tab:dbscan} and increased total running time, as can be seen in Figure \ref{fig:dbscan-scaling}. We are exploring ways of better partitioning the initial data to resolve the load imbalance issue while maintaining the scalability with increasing number of dimensions.

\begin{table}
\caption{\small Load distribution from geometric partitioning in DBSCAN}
\label{tab:dbscan}
\begin{tabular}{|l | r r r r|}
\hline
Nodes & 128 & 256 & 512 & 1024 \\ 
\hline
Min & 87392 & 81960	& 84963	& 70824 \\
Max & 130867	& 130147	& 154574	& 172377 \\
Average & 105156	& 105156	& 105156	& 105156 \\
Median & 104378	& 104759	& 103854	& 103666 \\
\hline 
\end{tabular}
\end{table}

For strong scaling on Haswell and KNL, we used a single timestep of the 1152 x 768 spatial field per node for the 128-nodes run; one timestep across 2 nodes for the 256-nodes run; one timestep across 4 nodes for the 512-nodes run; and one timestep across 8 nodes for the 1024-nodes run. We achieved an overall scaling efficiency of 38\% and clustering efficiency of 52\% on 1024 Haswell nodes. Increasing the number of nodes, while preserving the total input data size, results in a proportional decrease of partition sizes gathered per node. From the quadratic dependency on the number of points mentioned earlier, reducing the sizes of the partitions by $2x$, decreases the execution time by $4x$. However, since the partitions are not balanced, the obtained efficiency from increasing the number of nodes is marginally lower than the expected $4x$ reduction in execution time.

\subsection{Hero Run}
Our hero run processed 89.5 TB of lightcone data (obtained from 580 GB of simulated climate data) with distributed K-Means clustering on 1024 Intel Haswell nodes with 2 MPI ranks/node and 32 tbb threads/processor. We do not use Cori burst buffer. 580GB of climate data is read from the Cori /cscratch/ filesystem for generating nearly 90TB of lightcone data, after \textit{extract}, which resides in the on-node memory. The left column of Figure \ref{fig:kmeans-scaling} shows execution times for this run. The total time to solution was 6.6 minutes with a weak scaling efficiency of 91\%, which suggests that further scaling may be possible to process unprecedented amounts of scientific data and facilitate physics-based discovery in realistic systems.

\begin{figure*}
\centering
\includegraphics[width=0.9\textwidth,keepaspectratio]{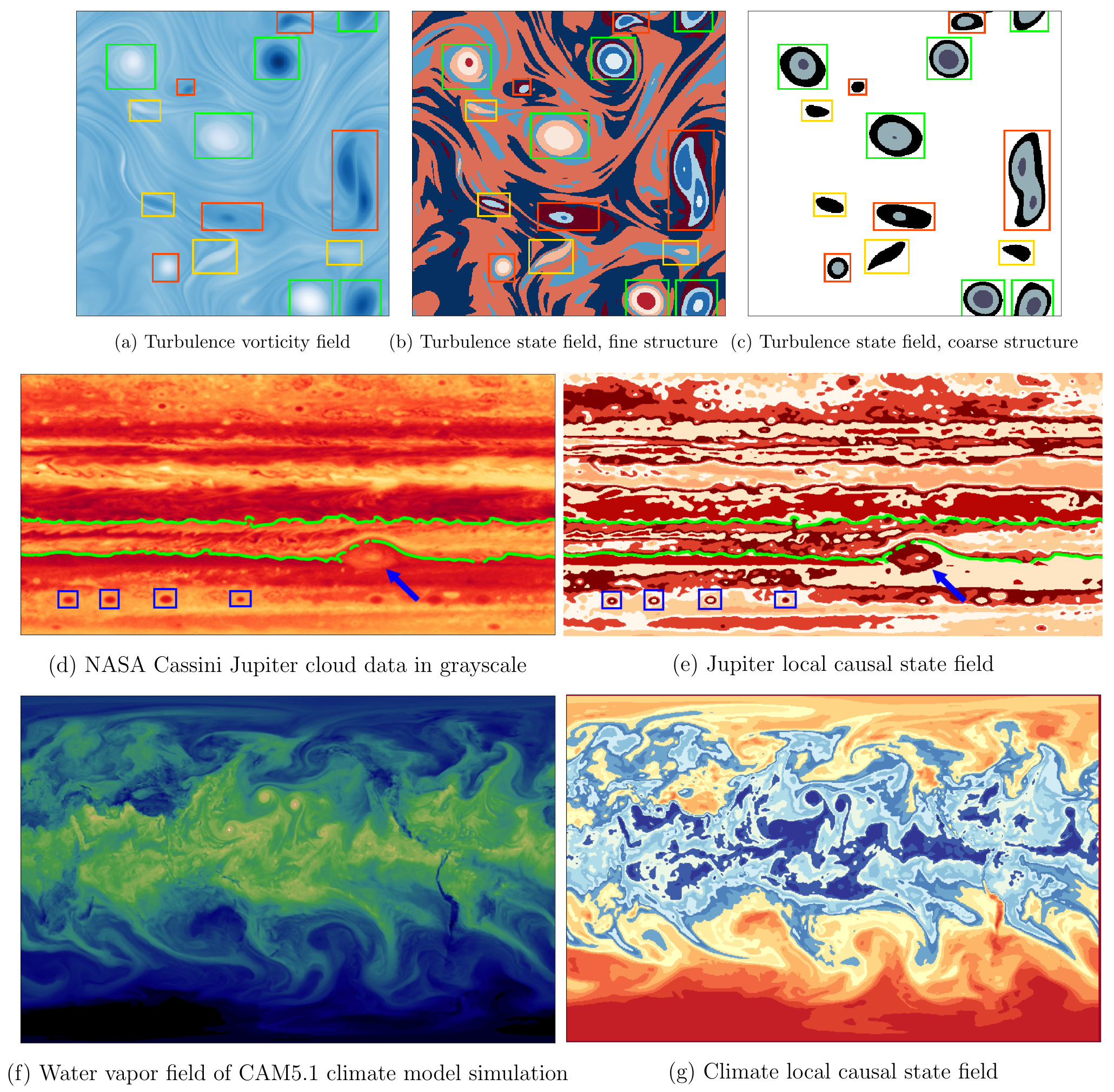}
% \vspace{-7mm}
\caption{\small Structural segmentation results for the three scientific data sets using K-Means lightcone clustering. The leftmost image of each row shows a snapshot from the data spacetime fields, and the other image(s) in the row show corresponding snapshots from the reconstructed local causal state spacetime fields. Reconstruction parameters given as $(h^-, h^+, c, K^-, \tau)$: (b) - (14, 2, 1, 10, 0.8), (c) - (14, 2, 1, 4, 0.0), (e) - (3, 3, 3, 8, 0), (g) - (3, 3, 1, 16, 0.04). $K^+ = 10$ and $0.05$ for chi-squared significance level were used for all reconstructions. 
%The left column contains sample images from the data spacetime fields and the right shows the corresponding sample image from the local causal state spacetime fields that are the outputs of the DisCo reconstruction pipeline. For all three reconstructions lightcone parameters $(h^-, h^+, c) = (3, 3, 1)$ were used. For the turbulence and Jupiter reconstructions $K=8$ for $\gamma^-$ and $K=10$ for $\gamma^+$. For climate, $K=16$ for $\gamma^-$ and $K=10$ for $\gamma^+=10$. 
Full segmentation videos are available on the DisCo YouTube channel \cite{discoyoutube}}
\label{fig:science}
\end{figure*}

\section{Science Results}
\label{scienceresults}

Snapshot images for our segmentation results on the three scientific data sets using K-Means clustering in the DisCo pipeline are shown in Figure~\ref{fig:science}.
%Sample segmentation images for the three data sets using K-Means clustering in the DisCo pipeline are shown in Figure~\ref{fig:science}. 
DBSCAN results are discussed at the end of this section. We emphasize that DisCo produces a \emph{spacetime} segmentation; the images shown are single-time snapshots taken from spacetime videos. The left image of each row in Figure~\ref{fig:science} -- (a), (d), and (f) -- are snapshots from the unlabeled ``training'' data used for local causal state reconstruction. %As an unsupervised method, nothing else is needed beyond the raw spacetime data. 
The other image(s) in each row are corresponding snapshots from the local causal state segmentation fields.
%, which are the output of Alg. 5: causal filtering, the final stage in the DisCo pipeline. 
%Parameters used for reconstruction are given in the caption of Figure~\ref{fig:science}. 
Full segmentation videos are available at the DisCo YouTube channel \cite{discoyoutube}.
%Recall that for segmentation each point in spacetime (pixel in the video) is assigned a class label. 

%The segmentation masks shown in \cite{kurt18a} have the following semantics: \texttt{cyclone}, \texttt{atmospheric river}, and \texttt{null}, where \texttt{null} is assigned to background points that are neither cyclones nor atmospheric rivers. 
The extreme weather event segmentation masks shown in \cite{kurt18a} have the following semantics: \texttt{cyclone}, \texttt{atmospheric river}, and \texttt{background}. 
In contrast, the segmentation classes of DisCo are the local causal states. Each unique color in the segmentation images -- Figure~\ref{fig:science} (b), (c), (e), and (g) -- corresponds to a unique local causal state. Further post-processing is needed to assign semantic labels such as \texttt{cyclone} and \texttt{atmospheric river} to sets of local causal states. We will discuss this further in Sec.~\ref{scienceresults}.1 and Sec.~\ref{scienceresults}.3. 

\subsubsection{Structural Decomposition}
The local causal state fields that are the direct output of DisCo, without additional semantic labels, can be considered a ``structural decomposition'' of the flow.
%DisCo incorporates the physics of local interactions across space and time to produce a local causal state segmentation that decomposes the flow into coherent spacetime sets. 
Incorporating the physics of local interactions to generalize the computational mechanics theory of structure to spatiotemporal systems, the local causal states are a more principled and well-motivated decomposition approach compared to empirical dimensionality reduction methods such as PCA and DMD (see Sec.~\ref{sec:RelWork}), or automated heuristics like TECA. 

But does the structural decomposition of the local causal states capture meaningful ``structure''?
What constitutes physically meaningful structure in complex fluid flows is an incredibly challenging open problem \cite{Holm12a, Hadj17a}. Even something as seemingly obvious as a fluid vortex does not have a generally accepted rigorous definition \cite{Epps17a}. 
%This is to say that there currently is no ground truth for the fluid structures we are interested in, and so it is impossible to give a quantitative assessment of how close our method gets to ground truth.  
This is to say that it is impossible to give a quantitative assessment of how close our method gets to ground truth because ground truth for this problem currently does not exist. 

\subsubsection{Lagrangian Coherent Structures}
In the absence of a quality metric to compare different methods against, the community standard is to qualitatively compare methods against each other. In particular, the Lagrangian approach to coherent structures in complex fluids is gaining wide acceptance and \cite{Hadj17a} surveys the current state-of-the-art Lagrangian Coherent Structure methods (see Sec.~\ref{sec:RelWork}). We directly compare our results with the geodesic and LAVD approaches (described below) on the 2D turbulence data set from \cite{Hadj17a} and the Jupiter data set from \cite{Hadj17a} and \cite{Hadj16a}.  

%In the LCS framework, there are generally three classes of structures that methods attempt to identify: elliptic, parabolic, and hyperbolic LCS \cite{Hall15a}. 
There are three classes of flow structures in the LCS framework;
elliptic LCS are rotating vortex-like structures, parabolic LCS are generalized Lagrangian jet-cores, and hyperbolic LCS are tendril-like stable-unstable manifolds in the flow. 
The geodesic approach \cite{Hall15a, Hadj16a} is the state-of-the-art method designed to capture all three classes of LCS and has a nice interpretation for the structures it captures in terms of characteristic deformations of material surfaces. The Lagrangian-Averaged Vorticity Deviation (LAVD) \cite{Hall16a} is the state-of-the-art method specifically for elliptic LCS, but is not designed to capture parabolic or hyperbolic LCS. 

The local causal states are not a Lagrangian method (they are built from spacetime fields, not Lagrangian particle flow) so they are not specifically designed to capture these structures. However, LCS are of interest because they heavily dictate the dynamics of the overall flow, and so signatures of LCS should be captured by the local causal states. As we will see in the results and comparisons in 7.1 and 7.2, this is indeed the case.

\subsubsection{Reconstruction Parameters}
%Before we move to the LCS results, a discussion of how reconstruction parameters affect the local causal state structural decomposition is in order.
The complex fluid flows of interest are multi-scale phenomena and so the question of how they are structured may not have a single answer. Different notions of structure may exist at different length and time scales. 
With this in mind, we have found that essentially all reconstruction parameters yield a physically valid structural decomposition. Varying parameters adjusts the structural details captured in a satisfying way.
%With this in mind, we have found from experiments on these data sets that the parameters for local causal state reconstruction do not need to be tuned just right to identify physically valid structure. Rather, the resulting structural decomposition varies in a satisfying way as reconstruction parameters are varied. 

Larger values of $K$ in K-Means produce refinements of structural decompositions from smaller values of $K$, capturing finer levels of detail. The speed of information propagation $c$ controls the spatial-scale of the structural decomposition and the decay-rate $\tau$ controls the temporal coherence scale. Because uniqueness and optimality of local causal states are asymptotic properties, lightcone horizons should be set as large as is computationally feasible. Though some finite cutoff must always be used. The lightcone horizon creates a discrete cut in the amount information of local pasts that is taken into account, as opposed to the smooth drop-off of the temporal decay. 

The local causal states, using different parameter values, provide a suite of tools for analyzing structure in complex, multi-scale spatiotemporal systems at various levels of description. Finally, we note that the $\tau \rightarrow \infty$ (or, equivalently the $h^\pm \rightarrow 0$) limit produces a standard K-Means image segmentation, which captures instantaneous structure and does not account for coherence across time and space.

\subsection{2D Turbulence}
% Relating DisCo's local causal state decomposition results with LCS in this way is cleanest with the idealized 2D turbulence data: Figure~\ref{fig:science} (a) and (b). The largest magnitude (darkest color) vorticity regions in Figure~\ref{fig:science} (a) are identified as elliptic LCS; see Figure 9 in \cite{Hadj17a}, particularly (k) and (l). Each of these elliptic LCS is generally captured by two concentric local causal states in Figure~\ref{fig:science} (b); one state for the high-rotation core and one for the boundary region. Hyperbolic LCS are harder to discern from these static images, but they generally form between the rotating elliptic LCS. They can be identified in the local causal state field as, for example, narrow white bands between blue and red states emanating from elliptic LCS, visualized more clearly in the videos on the YouTube channel \cite{discoyoutube}. 
While still complex and multi-scale, the idealized 2D turbulence data provides the cleanest Lagrangian Coherent Structure analysis using our DisCo structural decomposition. Figure~\ref{fig:science} (a) shows a snapshot of the vorticity field, and (b) and (c) show corresponding snapshots from structural decompositions using different reconstruction parameter values. Both use the same lightcone template with $h^-=14$, $h^+=2$, and $c=1$. To reveal finer structural details that persist on shorter time scales, Figure~\ref{fig:science} (b) uses $\tau=0.8$ and $K=10$. To isolate the coherent vortices, which persist at longer time scales, Figure~\ref{fig:science} (c) was produced using $\tau=0.0$ and $K=4$. As can be seen in (b), the local causal states distinguish between positive and negative vortices, so for (c) we modded out this symmetry by reconstructing from the absolute value of vorticity. 

All three images are annotated with color-coded bounding boxes outlining elliptic LCS to directly compare with the geodesic and LAVD LCS results from Figure 9, (k) and (l) respectively, in \cite{Hadj17a}. Green boxes are vortices identified by both the geodesic and LAVD methods and red boxes are additional vortices identified by LAVD but not the geodesic. Yellow boxes are new structural signatures of elliptic LCS discovered by DisCo. Because there is a single background state, colored white, in (c), all objects with a bounding box can be assigned a semantic label of \texttt{coherent structure} since they satisfy the local causal state definition given in \cite{Rupe18b} as spatially localized, temporally persistent deviations from generalized spacetime symmetries (i.e. local causal state symmetries). 
%That is, DisCo provides the physical interpretation of coherent structures as locally broken symmetries, and in (c) we discover vortices as coherent structures in this very general sense.
Significantly, DisCo has discovered vortices in (c) as coherent structures with this very general interpretation as locally broken symmetries. 

In the finer-scale structural decomposition of (b) we still have a unique set of states outlining the coherent vortices, as we would expect. If they persist on longer time scales, they will also persist on the short time scale. The additional structure of the background potential flow largely follows the hyperbolic LCS stable-unstable manifolds. Because they act as transport barriers, they partition the flow on either side and these partitions are given by two distinct local causal states with the boundary between them running along the hyperbolic LCS in the unstable direction. For example, the narrow dark blue-colored state in the upper right of (b) indicates a narrow flow channel squeezed between two hyperbolic LCS. 

\subsection{Jupiter}
% For Jupiter, Figure~\ref{fig:science} (c), the Great Red Spot is the most obvious elliptic LCS; see \cite{Hadj17a} Figure 14 and \cite{Hadj16a}. But there are also several smaller vortices that are captured well by DisCo's local causal states; center left in Figure~\ref{fig:science} (d) are the easiest to see. The opposing east-west flowing cloud bands of Jupiter make it easier to see hyperbolic LCS from static images; the shear layer boundaries that form between neighboring bands act as a transportation barrier in the north-south direction. This is reflected in Figure~\ref{fig:science} (d) as two distinct local causal states straddling shear layer boundaries, extending in the east-west direction. 

%The graininess of the Jupiter segmentation is likely due to short flow duration available from the Cassini video and the small lightcones used in reconstruction. 

Figure~\ref{fig:science} (d) shows a snapshot from the Jupiter cloud data, with corresponding structural decomposition snapshot in (e). The Great Red Spot, highlighted with a blue arrow, is the most famous structure in Jupiter's atmosphere. As it is a giant vortex, the Great Red Spot is identified as an elliptic LCS by both the geodesic and LAVD methods \cite{Hadj16a, Hadj17a}. While the local causal states in (e) do not capture the Great Red Spot as cleanly as the vortices in (b) and (c), it does have the same nested state structures as the turbulence vortices.
%Recall that the Jupiter data is cloud luminosity and not fluid vorticity. 
There are other smaller vortices in Jupiter's atmosphere, most notably the ``string of pearls'' in the Southern Temperate Belt, four of which are highlighted with blue bounding boxes. We can see in (e) that the pearls are nicely captured by the local causal states, similar to the turbulence vortices in (b). 

Perhaps the most distinctive features of Jupiter's atmosphere are the zonal belts. The east-west zonal jet streams that form the boundaries between bands are of particular relevance to Lagrangian Coherent Structure analysis. Figure 11 in \cite{Hadj16a} uses the geodesic method to identify these jet streams as shearless parabolic LCS, indicating they act as transport barriers that separate the zonal belts.
%These east-west flow bands are highly structured, in a general sense, but their most significant features relevant to Lagrangian Coherent Structures are the parabolic LCS transport barriers that form the boundaries between bands. Figure 11 in \cite{Hadj16a} shows shearless parabolic LCS that form the core of the zonal jet streams, as identified by the geodesic method. 
%Transport barriers appear as boundaries between distinct local causal states in the DisCo structural decomposition.
The particular decomposition shown in (e) captures a fair amount of detail inside the bands, but the edges of the bands have neighboring pairs of local causal states with boundaries that extend contiguously in the east-west direction along the parabolic LCS transport barriers. Two such local causal state boundaries are highlighted in green, for comparison with Figure 11 (a) in \cite{Hadj16a}. The topmost green line, in the center of (d) and (e), is the southern equatorial jet, shown in more detail in Figure 11 (b) and Figure 12 of \cite{Hadj16a}. Its north-south meandering is clearly captured by the local causal states. 

%Beyond the vortices and zonal band boundaries captured by the geodesic method as elliptic and parabolic LCS respectively, the local causal states decompose the full atmospheric flow and capture structure in a more general sense. 

%In addition to the vortices (elliptic LCS) and zonal band boundaries (parabolic LCS), the local causal states decompose the full atmospheric flow and capture its structure in a more general sense.

\subsection{Extreme Weather Events}
Finally, we return to climate science. Figure~\ref{fig:science} (g) shows the local causal state decomposition of the CAM5.1 water vapor field shown in Figure~\ref{fig:science} (f). While our decomposition of the turbulence and Jupiter data align nicely with the LCS analysis in \cite{Hadj17a} and \cite{Hadj16a}, we are not yet able to use the climate decomposition to construct segmentation masks of weather patterns.

However, given the promising decomposition in Figure~\ref{fig:science} (g), we believe this is achievable. 
%\cite{Rupe18b} demonstrated, with cellular automata, that a principled coherent structure semantic analysis on top of local causal state decomposition is possible. Furthermore, 
The climate decomposition shown here was performed solely on the water vapor field, and not all physical variables of the CAM5.1 climate model, like was done in \cite{kurt18a}. While we see signatures of cyclones and atmospheric rivers outlined in Figure~\ref{fig:science} (g), it is not surprising that these structures are not uniquely identified from the water vapor fields alone; this would be akin to describing hurricanes as just local concentrations of water vapor. More contextual spacetime information is needed. This includes additional physical fields, and the use of larger lightcones in reconstruction. 

A key distinguishing feature of hurricanes is their rotation. While rotation has its signature in the water vapor field, the three timesteps used for the lightcones in our local causal state reconstruction cannot capture much of this rotation. The vorticity field, as used for the 2D turbulence data, gives instantaneous local circulation. From Figure~\ref{fig:science} (b) and (c) we see that vortices are easier to capture from vorticity. Additionally, hurricanes have distinctive patterns in temperature and pressure, for example a warm, low pressure inner core. Including the vorticity, temperature, and pressure fields into the climate decomposition will help yield a distinct set of hurricane states. 

Similarly, atmospheric rivers (ARs) have distinguishing characteristics in other physical fields, most notably water vapor transport, that will help identify ARs when added to the decomposition. The use of larger lightcones should be particularly helpful for creating AR masks, as their geometry is crucial for their identification, which extends over a much larger spatial scale than can be captured by the depth-3 lightcones used in the reconstruction here.

\begin{figure*}
\centering
\includegraphics[width=0.8\textwidth,keepaspectratio]{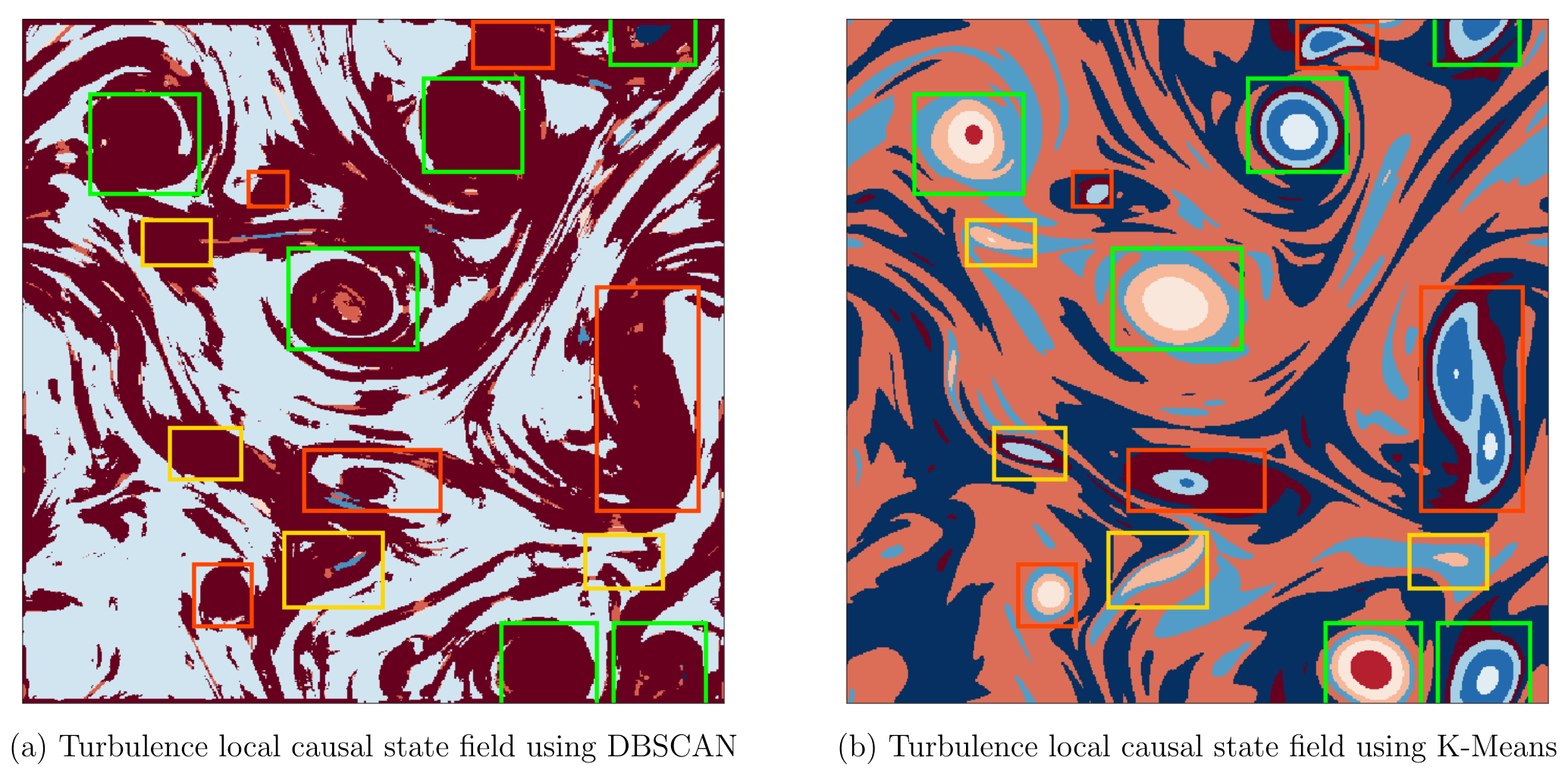}
% \vspace{-7mm}
\caption{\small Comparison of structural segmentation results on 2D turbulence using DBSCAN (a) and K-Means (b) for lightcone clustering. The K-Means results in (b) are the same as Figure~\ref{fig:science} (b), repeated here for easier comparison. The DBSCAN results shown in (a) use reconstruction parameters $(h^-, h^+, c) = (3,2,1)$, $\tau = 0.0$, \texttt{eps} = $0.0$, and \texttt{minpts} = $10$.}
\label{fig:dbscan}
\end{figure*}

\subsection{Lightcone Clustering Revisited}
%So what about DBSCAN, which was expected to be the more appropriate clustering method for the DisCo pipeline? 
% For DBSCAN we found that most typical outcomes either classify most points as noise or into one single cluster. Some density parameters give $O(1000)$ clusters, but most contain $O(1)$ lightcones. All cases do not yield meaningful segmentation output. 

As a density-based method, when compared to K-Means, DBSCAN faces the curse of dimensionality \cite{Zime12a, Mari79a}. Distinguishing density-separated regions becomes exponentially difficult in high-dimensional spaces, such as the lightcone spaces used in our applications. Further, a limiting assumption of DBSCAN is that all clusters are defined by a single density threshold. Adaptive methods like OPTICS \cite{Anke99a} and HDBSCAN \cite{hdbscan} may perform better for complex data. The results we observe from experiments with DBSCAN suggests that the lightcone-spaces of complex fluid flows do not contain clear density-separated regions, and thus do not yield to a density-based discretization. 
%This would be consistent with the smoothness and continuity of the flows. 

Figure~\ref{fig:dbscan} shows snapshots of spacetime segmentation results of the turbulence data set using both DBSCAN (a) and K-Means (b). The K-Means result in Figure~\ref{fig:dbscan} (b) is copied from Figure~\ref{fig:science} (b) for easier visual comparison with the DBSCAN results in Figure~\ref{fig:dbscan} (a), which used reconstruction parameters $(h^-, h^+, c) = (3,2,1)$, $\tau=0.0$, \texttt{eps}=$0.0005$, and \texttt{minpts} = 10. We can see that the DBSCAN segmentation does outline structure in the flow, but it is all detailed structure of the background potential flow. None of our experiments with different parameter values were able to isolate vortices with a unique set of local causal states, which, as demonstrated above, is possible with K-Means. This inability to isolate vortices, along with the patchwork pattern of decomposition in parts of the background flow suggest that a single density threshold of lightcone space, which is assumed by DBSCAN, is incapable of coherent structure segmentation. 

Similarly, our DBSCAN experiments on the climate data typically yielded outcomes that either classify most points as noise or into one single cluster. Some density parameters give $O(1000)$ clusters, but most contain $O(1)$ lightcones. All of these cases do not yield physically meaningful segmentation output, which again points to the shortcoming of a single density threshold. As can be seen in Figure~\ref{fig:science} (f), the CAM5.1 water vapor data is very heterogeneous in space; the water vapor values are much more uniform towards the poles (the top and bottom of the image) than in the tropics (center). The polar regions will contribute a relatively small, but very dense, region in lightcone space compared to contributions from the tropics.

From experiments it appears that K-Means attempts to uniformly partition lightcone-space, which is consistent with these spaces not being density-separated. If this is in fact the case, the convex clusters that uniformly separate lightcone-space which result from K-Means are actually the most natural discretization of lightcone-space concordant with the continuous histories assumption.

Despite prior assumptions and intuitions, K-Means appears to be a much more effective clustering method for hydrodynamic coherent structure segmentation using our DisCo pipeline. That being said, there are plenty of other applications for which density-based clustering using DBSCAN is the appropriate choice. Our DBSCAN implementation has now made this algorithm available for large, high-dimensional applications, with the same easy-to-use Python API as found in scikit-learn.

\section{Conclusions}
\label{sec:conclusions}

We present several firsts in this paper, including a physics-based behavior-driven approach for unsupervised segmentation of coherent spatiotemporal structures in scientific data. Our high-performance distributed pipeline, written entirely in Python, with newly developed libraries, demonstrates the application of K-Means and DBSCAN clustering methods to $\sim 100$ dimensional clustering problems with over $70B$ points. We achieve 91\% weak and 64\% strong scaling efficiency with distributed K-Means clustering on 1024 Intel\registered Haswell nodes (32-core). We also demonstrate state-of-the-art segmentation results for three complex scientific datasets, including on 89.5 TB of simulated climate data (lightcones) in 6.6 minutes end-to-end. Our pipeline can be scaled further, which will facilitate accurate and precise structural segmentation on unprecedented amounts of scientific spatiotemporal data.
\section*{Disclaimers}
\footnotesize
\noindent Software and workloads used in performance tests may have been optimized for performance only on Intel microprocessors. 

\noindent Optimization Notice: Intel's compilers may or may not optimize to the same degree for non-Intel microprocessors for optimizations that are not unique to Intel microprocessors. These optimizations include SSE2, SSE3, and SSSE3 instruction sets and other optimizations. Intel does not guarantee the availability, functionality, or effectiveness of any optimization on microprocessors not manufactured by Intel. Microprocessor-dependent optimizations in this product are intended for use with Intel microprocessors. Certain optimizations not specific to Intel microarchitecture are reserved for Intel microprocessors. Please refer to the applicable product User and Reference Guides for more information regarding the specific instruction sets covered by this notice. Notice Revision \#20110804

\noindent Performance tests, such as SYSmark and MobileMark, are measured using specific computer systems, components, software, operations and functions. Any change to any of those factors may cause the results to vary. You should consult other information and performance tests to assist you in fully evaluating your contemplated purchases, including the performance of that product when combined with other products. For more information go to www.intel.com/benchmarks.

\noindent Performance results are based on testing as of April 10, 2019 and may not reflect all publicly available security updates. See configuration disclosure for details.  No product or component can be absolutely secure.

\noindent Configurations: Testing on Cori (see \ref{sec:setup}) was performed by NERSC, UC Davis, and Intel, with the spectre\_v1 and meltdown patches.

\noindent Intel technologies' features and benefits depend on system configuration and may require enabled hardware, software or service activation. Performance varies depending on system configuration. Check with your system manufacturer or retailer or learn more at [intel.com].

\noindent Intel, the Intel logo, Intel Xeon, Intel Xeon Phi, Intel DAAL are trademarks of Intel Corporation or its subsidiaries in the U.S. and/or other countries. *Other names and brands may be claimed as the property of others. 

\section*{Acknowledgements}
AR and JPC would like to acknowledge
Intel\registered for supporting the IPCC at UC Davis.
KK and P were supported by the
Intel\registered Big Data Center. This research is based upon work
supported by, or in part by, the U. S. Army Research
Laboratory and the U. S. Army Research Office under
contracts W911NF-13-1-0390 and W911NF-18-1-0028, and used resources of the
National Energy Research Scientific Computing Center,
a DOE Office of Science User Facility supported by the
Office of Science of the U.S. Department of Energy
under Contract No. DE-AC02-05CH11231.

%%%%%%%%%%%%%%%%%%%%%%%%%%%%%%%%%%%%%%%%%%%%%%%%%%%%%%%%%%%%%%%%%%%%%%%%%%%%

\bibliographystyle{IEEEtran}
\bibliography{disco}

% \bibliography{disco.bib}
% \end{thebibliography}

%%%%%%%%%%%%%%%%%%%%%%%%%%%%%%%%%%%%%%%%%%%%%%%%%%%%%%%%%%%%%%%%%%%%%%%%%%%%

\end{document}